\documentclass[manuscript]{aastex}

\shorttitle{Numerical Simulations of Chromospheric Microflares}
\shortauthors{Jiang, R. L. et al.}

\begin{document}

\title{Numerical Simulations of Chromospheric Microflares}

\author{R. L. Jiang, C. Fang and P. F. Chen}

\affil{Department of Astronomy, Nanjing University, Nanjing 210093, China}
\email{Email: fangc@nju.edu.cn}

\begin{abstract}

With gravity, ionization, and radiation being considered,
we perform 2.5D compressible resistive MHD
simulations of chromospheric magnetic reconnection using the
CIP-MOCCT scheme. The temperature distribution of the
quiet-Sun atmospheric model VALC and the helium abundance ($10\%$)
are adopted. Our 2.5D MHD simulation reproduces
qualitatively the temperature enhancement observed in chromospheric
microflares. The temperature enhancement $\Delta T$ is demonstrated
to be sensitive to the background magnetic field, whereas the total
evolution time $\Delta t$ is sensitive to the magnitude of the
anomalous resistivity. Moveover, we found a scaling
law, which is described as $\Delta T/\Delta t \sim {n_H}^{-1.5}
B^{2.1} {\eta_0}^{0.88}$. Our results also indicate that the
velocity of the upward jet is much greater than that of the
downward jet and the X-point may move up or down.
\end{abstract}


\section{INTRODUCTION}

Magnetic reconnection plays a very important role in solar flares,
corona mass ejections, and other solar activities. During the solar
minimum, many authors focus their attention on the solar small-scale
activities such as microflares \citep{Qiu2004, Fang2006a,
Ning2008, Brosius2009}, Ellerman bombs \citep{Fang2006b,
Watanabe2008}, chromospheric jets \citep{Nishizuka2008}, and so on.

Microflares, or subflares, or bright points, are small-scale and
short-lifetime solar activities. The size of microflares ranges from
several arcsecs to about 20 arcsecs, the duration and the total energy
can be $10-30$ minutes and $10^{26}-10^{29}$ ergs, respectively
\citep{Shimizu2002, Fang2006a}. The most distinctive feature in their
visible spectra is the faint emission at the center and the brightening
at the wing of some chromospheric lines such as H$\alpha$ line. Soft
X-ray \citep{Golub1974, Golub1977}, hard X-ray \citep[HXR][]{Lin1984},
EUV \citep{Porter1984, Emslie1978} and microwave \citep{Gary1988}
emissions have also been observed in some microflares. It is an interesting fact that there are some correlations among the
emissions at different wavelengths. For instance, \cite{Qiu2004} found
that about 40\% of microflares show HXR emissions at $>$10 keV and
microwave emissions at $\sim$10 GHz. Recently, \cite{Ning2008} found
that there is a correlation between the power-law index of the HXR
spectrum and the emission measure of microflares. \cite{Brosius2009}
found that the studied microflare is bright first in the chromospheric
and transition region spectral lines rather than in the corona, which is consistent with the chromospheric heating by nonthermal electron beams. All these evidence indicates that microflares are related to the nonthermal processes driven by magnetic reconnection.

Some microflares might result from the magnetic reconnection in the
corona. However, there is accumulating evidence suggesting
that some other microflares are due to the reconnection in the
chromosphere \citep{Liu2004, Xia2007}. For instance, \cite{Tang2000}
found that in many cases emerging flux occurred about $5-30$ minutes
before the microflares, where the emerging flux may collide with the
pre-existing magnetic field at the chromospheric height.
\cite{Brosius2009} also stated that magnetic reconnection in the
chromosphere could be a plausible mechanism for triggering the
chromospheric microflare.

\cite{Chen2001} made 2.5-dimensional (2.5D) numerical simulations of
chromospheric magnetic reconnection in order to study Ellerman bombs
and type II white-light flares. However, in the simulations, they
made some simplifications, such as the omissionof the gravity.
In this paper, in order to improve their results, we perform
2.5D magnetohydrodynamic (MHD) simulations using the CIP-MOCCT scheme,
with gravity being included, and try to make some comparisons with the
observations of chromospheric microflares. In the next section, the
numerical method is given in detail. Our results, which include
the dynamic process, parameter dependence, a scaling law and a
comparison with a semi-empirical model, are described in Section
\ref{NUMERICAL RESULTS}.  Discussion and summary are given in Section
\ref{Disc}.

\section{NUMERICAL METHOD}
\label{NUMERICAL METHOD}

\subsection{Basic Equations}
\label{Basic Equations}

In our simulation, we solve 2.5D compressible resistive MHD
equations with ionization, radiation, and uniform gravitational
field being considered. There is no thermal conduction in
our simulation since thermal conductivity is very small in the
chromosphere.  The MHD equations are given as follows:

\begin{equation}
\frac{\partial \rho}{\partial t} + \nabla \cdot (\rho \mathbf{v}) =
0 \,\, ,
 \label{MHD-1}
\end{equation}

\begin{equation}
\frac{\partial (\rho \mathbf{v})}{\partial t} + \nabla \cdot [( p
+\frac{{B}^2}{2\mu_0}) \mathbf{I} + \rho \mathbf{v} \mathbf{v} -
\frac{1}{\mu_0} \mathbf{B} \mathbf{B}] = \rho \mathbf{g} \,\, ,
\label{MHD-2}
\end{equation}

\begin{equation}
 \frac{\partial \mathbf{B}}{\partial t} + \nabla \cdot (\mathbf{v}
  \mathbf{B} - \mathbf{B} \mathbf{v}) = - \nabla \times
  (\frac{\eta}{\mu_0} \nabla \times \mathbf{B}) \,\, ,
  \label{MHD-3}
\end{equation}

\begin{equation}
\frac{\partial e}{\partial t} + \nabla \cdot [\mathbf{v} (e +
\frac{{B}^2}{2\mu_0} + p) - \frac{1}{\mu_0}\mathbf{B}(\mathbf{B}
\cdot \mathbf{v})+(\frac{\eta}{{\mu_0}^2} \nabla \times
\mathbf{B})\times \mathbf{B}] = \rho \mathbf{g} \cdot \mathbf{v} + H
- R_r \,\, ,
\label{MHD-4}
\end{equation}

\noindent
where eight independent variables are the density ($\rho$), velocity
($v_x$, $v_y$, $v_z$), magnetic field  ($B_{x}$, $B_{y}$, $B_{z}$), and
temperature ($T$). The thermal and ionization terms are included into
the total energy density $e= p/(r-1) + \rho v^2/2 + B^2/2 \mu_0 + \chi_H
n_e$, where $\chi_{H}$ is the ionization potential of hydrogen atoms and
$\mu_0$ is the vacuum permeability. Considering the helium abundance
($10\%$), we take $\rho = 1.4 m_{H} n_{H}$, where $n_{H}$ is the number
density of hydrogen. With partial ionization considered, the total gas
pressure is expressed as $p = (1.1n_{H} + n_e ) k_B T$, where $k_B$ is
the Boltzmann constant and the number density of electrons, $n_{e}$, is
deduced from the modified Saha and Boltzmann formula for a pure hydrogen
atmosphere \citep{Gan1990}:

\begin{equation}
n_{e}=(\sqrt{\phi^{2} + 4 n_{H} \phi} - \phi)/2 \,\, ,
 \label{Ne-1}
\end{equation}

\begin{equation}
\phi = \frac{1}{b_{1}} (\frac{2 \pi m_{e}\kappa_B
T}{h_{0}^{2}})^{3/2} e^{\frac{-\chi_{H}}{\kappa_B T}} \,\, ,
 \label{Ne-2}
\end{equation}

\begin{equation}
b_{1}=\frac{2T}{T_{R}} e^{\frac{\chi_{H}}{4 \kappa_B T}
(\frac{T}{T_{R}} - 1)} \,\, ,
 \label{Ne-3}
\end{equation}

\noindent where $h_0$ is the Planck constant. We take $T_{R}$ = 6000
K \citep{Brown1973}. In Equation (\ref{MHD-4}), $R_r$ and $H$ are
the radiative loss and the pre-heating term, respectively. Radiative
loss is important in the chromosphere and the photosphere. Generally
speaking, it should be calculated by solving simultaneously the
statistical equilibrium and radiative transfer equations along with
the MHD equations. However, it is not feasible to deal with it in our
2.5D numerical simulation. Instead, we fit the non-local thermodynamic
equilibrium (non-LTE) radiative losses in the VALC model for the
quiet-Sun atmosphere \citep{Vernazza1981} and the F1 semi-empirical
model for the weak flare \citep{Machado1980}, and then modify the
empirical formula given by \citet{Gan1990} in order to make it more
suitable for the small scale activities such as microflares:

\begin{equation}
R_r = n_{H} n_{e} \alpha (y) f(T) \,\, ,
 \label{Rad-1}
\end{equation}

\begin{equation}
\alpha(y) = 10^{(1.745 \times 10^{-3} y - 4.739)} + 8.0 \times 10^{-2}
 e^{- 3.701 \times 10^{-2} y } \,\, ,
\label{Rad-2}
\end{equation}

\begin{equation}
f(T) = 4.533 \times 10^{-23} (\frac{T}{10^{4}})^{2.874} \,\, ,
\label{Rad-3}
\end{equation}

\noindent where $\alpha (y)$ and $f(T)$ are functions of $y$ (the
height from $\tau_{5000} = 1$ in the photosphere) and the
temperature $T$, respectively. The unit of $y$ is km. The suitable
range of the height and the temperature in Equation
(\ref{Rad-1}) is $h < 2000$km and $T < 10^5$ K, respectively. The
pre-heating term is given by $H = n_{H}(n_{e} \alpha f)_{t = 0}$,
which balances the radiative loss before magnetic reconnection.

\subsection{Normalization and Parameters}
\label{Normalization and Parameters}

In order to non-dimensionalize the MHD Equations (\ref{MHD-1})
- (\ref{MHD-4}) in Section~\ref{Basic Equations}, the variables are
normalized by the quantities given in Table~\ref{Normalization Units}.
In this table, $m_H$ is the proton mass. The dimensionless MHD
Equations, which have a standard form of advection and non-advection
terms, are numerically solved with the CIP-MOCCT scheme (Kudoh 1999).
Besides the acoustic transit time scale $t_0$ used in the
nondimensionalization, another practical time scale is the Alfv\'en
transit time $\tau_{A} = L_{0} / v_A$, where $v_A$ is the Alfv\'en
velocity at the height of the reconnection X-point. In our simulations,
two cases are considered, where the magnetic reconnection X-point is
set at two different heights, i.e., 500 km and 1000 km. The
corresponding Alfv\'en velocities are 18.2 km s$^{-1}$ and 11.2
km s$^{-1}$, while the corresponding Alfv\'en transit time is 109.9 s
and 177.6 s, respectively.

\subsection{Initial and Boundary Conditions}
\label{Initial and boundary Conditions}

\begin{figure}
   \centering
   \includegraphics[width=300pt]{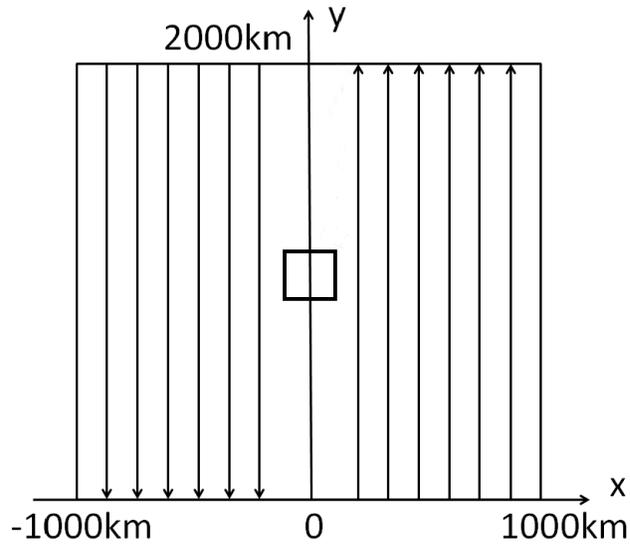}
   \caption{Initial magnetic field configuration. The range of
   the computational box is from -1000 km to 1000 km in the $x$
   direction and from 0 to 2000 km in the $y$ direction.
   \label{mag_field}}
\end{figure}

\begin{figure}
   \centering
   \includegraphics[width=150pt,height=150pt]{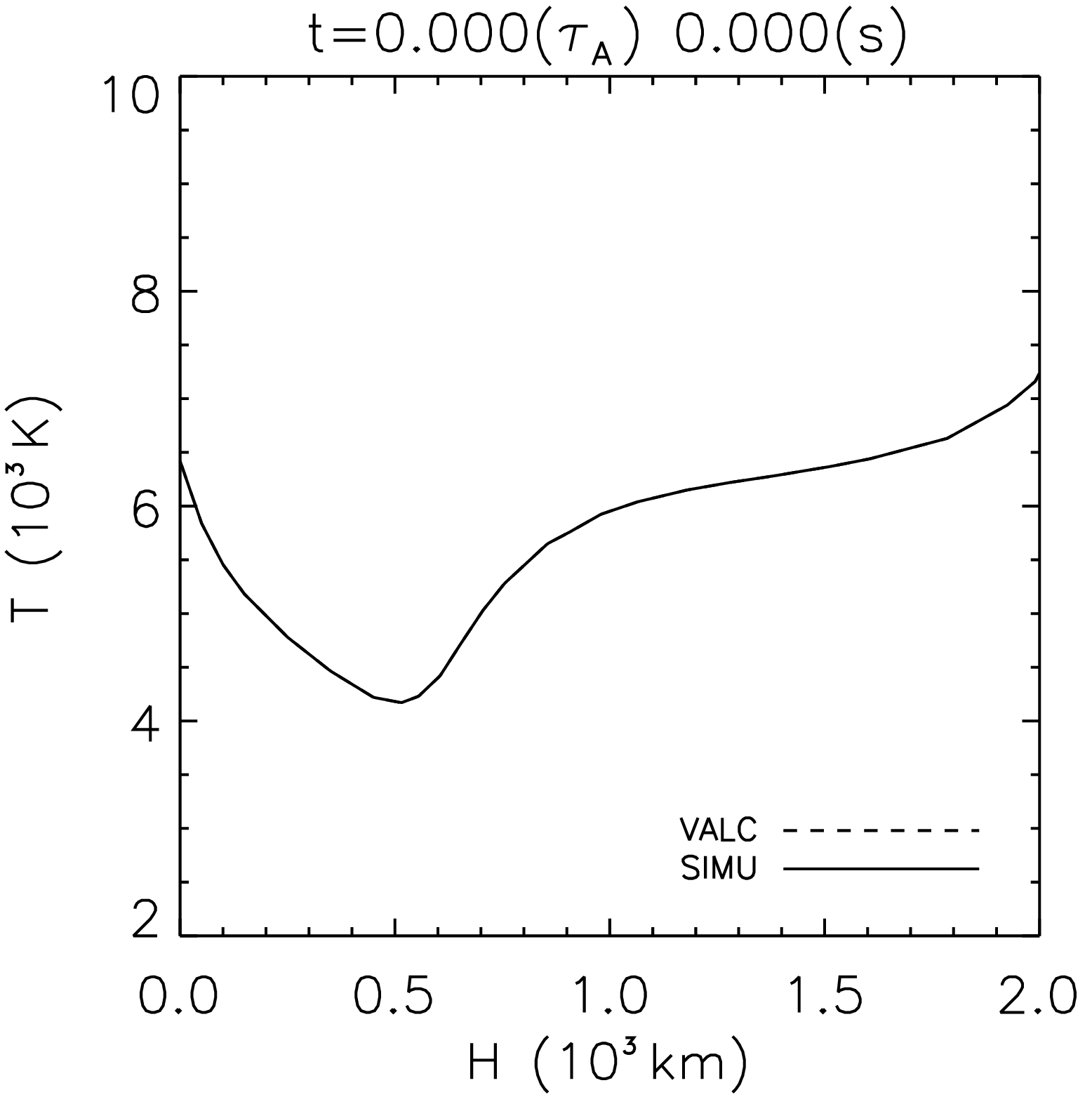}
   \includegraphics[width=150pt,height=150pt]{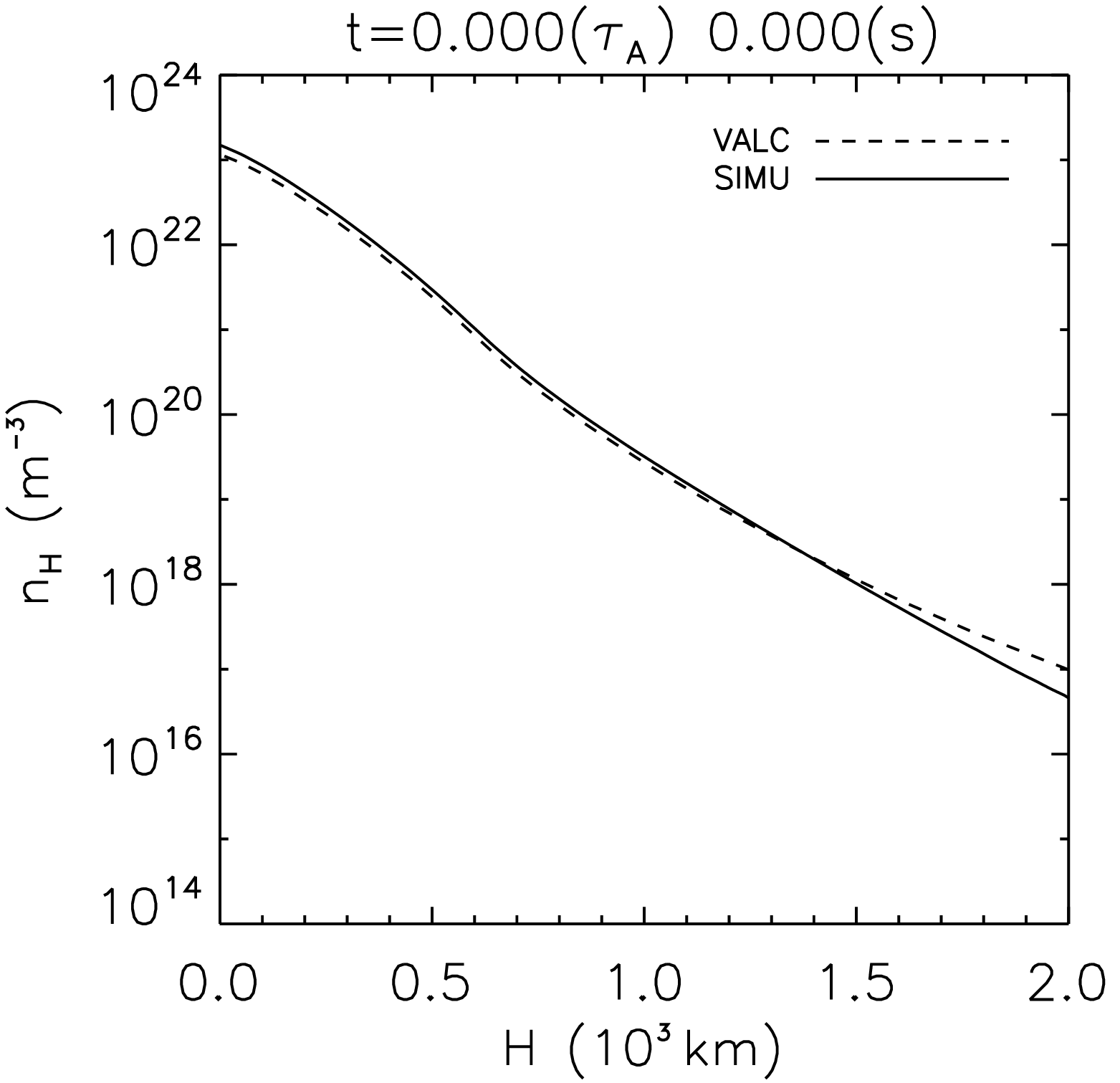}
   \includegraphics[width=150pt,height=150pt]{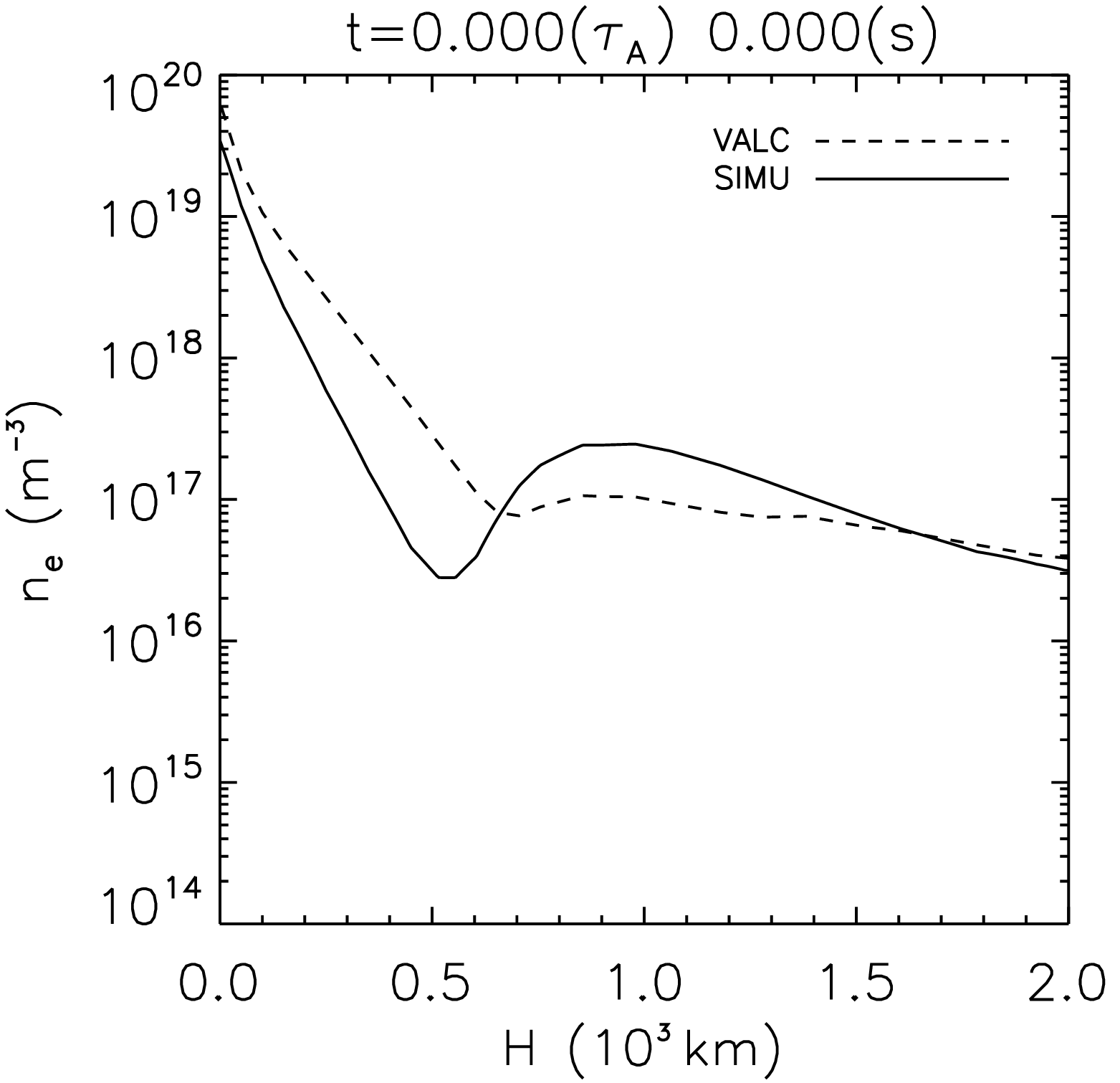}
    \caption{Initial distributions of the temperature (left
     panel), the number density of hydrogen (middle panel),
     and the electron number density (right panel). The
     distributions in the quiet-Sun atmospheric model VALC
     (dashed lines) are also shown.
   \label{fig_ini}}
\end{figure}

\begin{figure}
\begin{center}
\includegraphics[width=120pt,height=360pt]{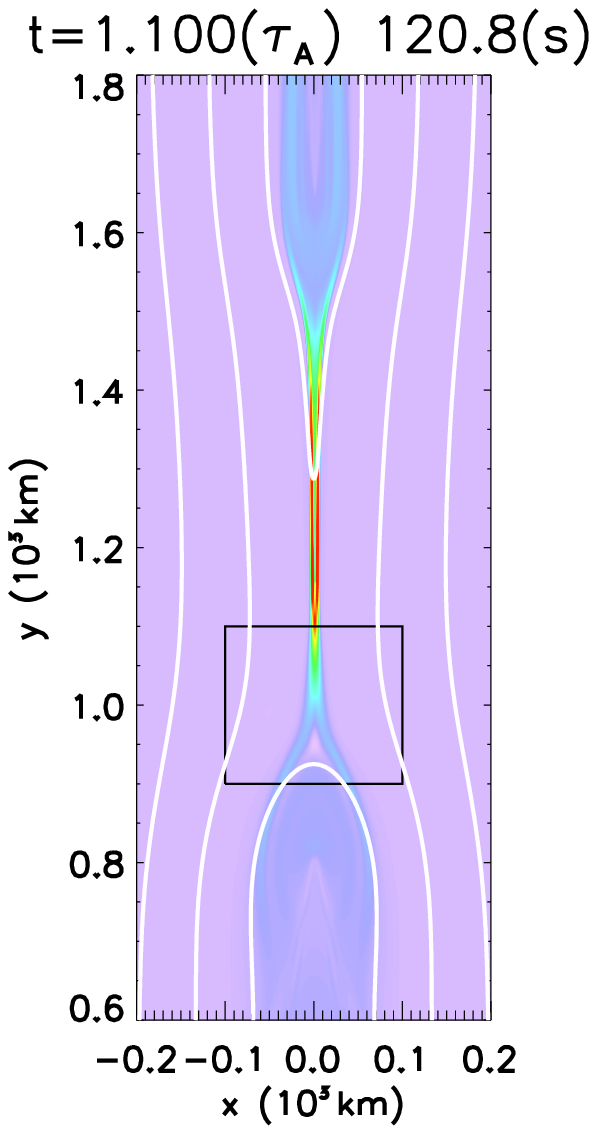}
\includegraphics[width=120pt,height=360pt]{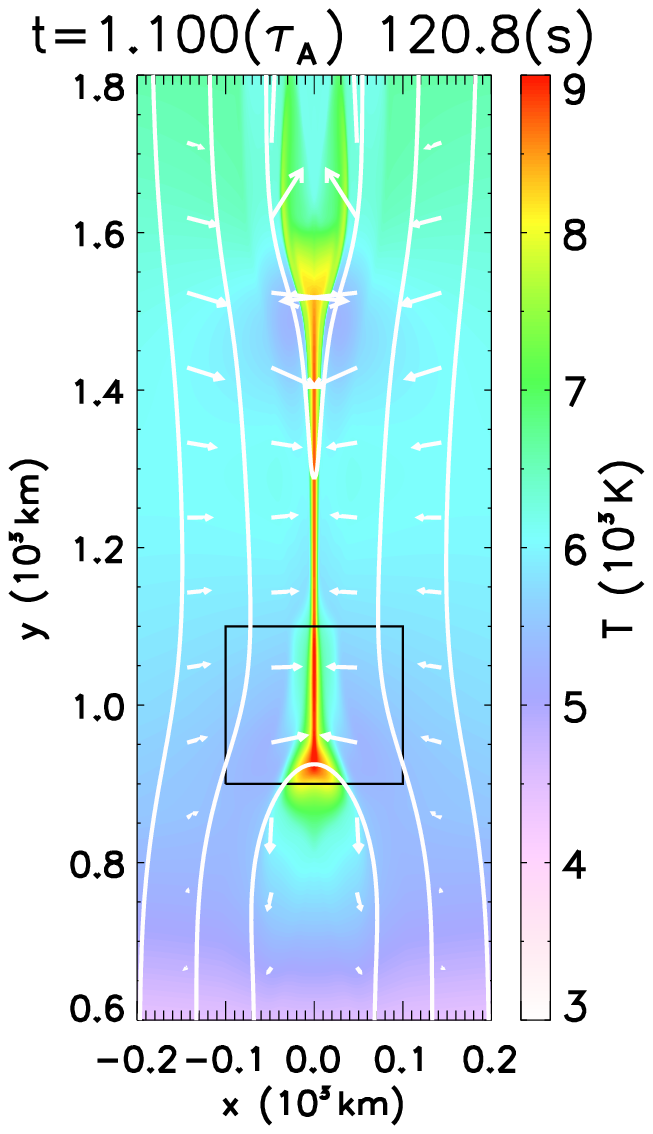}
\end{center}
\caption{Z-component of the current density (left panel) and
 temperature distributions (right panel) at the time of the maximum
 reconnection rate (1.1 $\tau_{A}$). There is no need to plot the
 color bar for the left panel which based on dimensionless data, but
 in both panels the same color table is used, which means that the
 red side stands for higher current density and temperature, while
 the white side stands for the lower ones.}
  \label{fig_sho}
\end{figure}

\begin{figure}
\begin{center}
\includegraphics[width=120pt,height=360pt]{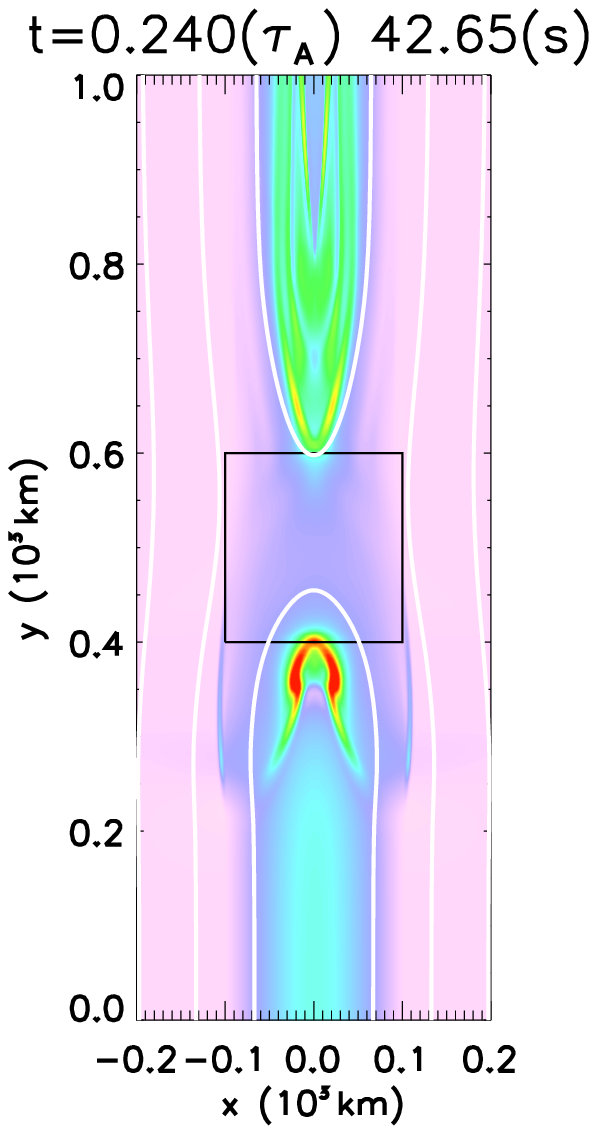}
\includegraphics[width=120pt,height=360pt]{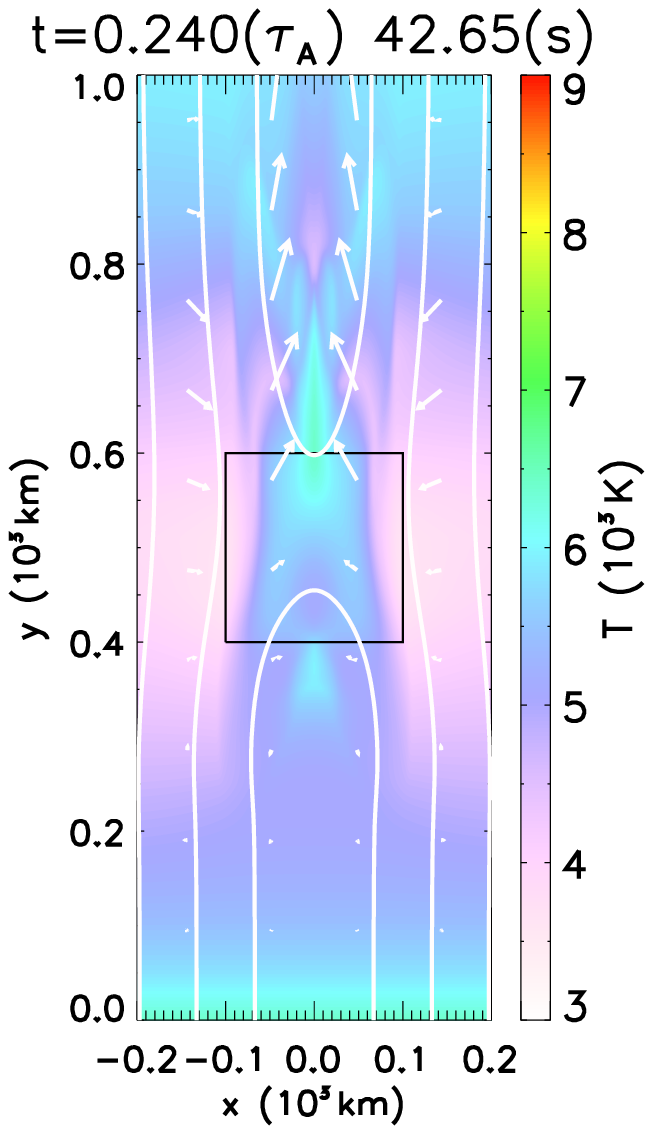}
\end{center}
\caption{Z-component of current density (left panel) and
 temperature distributions (right panel) at the time of maximum
 reconnection rate (0.24 $\tau_{A}$). Both panels have used the same
 color table, which means that the red side stands for higher current
 density and temperature, the white side stands for the lower ones.}
  \label{fig_Case_jz}
\end{figure}

The domain of the numerical simulation is -1000 km $\leq x \leq$ 1000
km, 0 $\leq y\leq$ 2000 km, as shown in Figure \ref{mag_field}. The
initial magnetic configuration is a force-free field with a current
sheet located along the $y$-axis, which was used by \cite{Chen1999}
and rewritten as follows:

\begin{equation}
  B_x = 0 \,\, ,
\end{equation}

\begin{equation}
  B_y=\left\{
  \begin{array}{ll}
    B_0\frac{x}{|x|}  & |x| > \triangle h  \\
    B_0 \sin{\frac{x \pi}{2 \triangle h }}  & |x| < \triangle h \,\,
    ,
  \end{array}
  \right.
\end{equation}

\begin{equation}
  B_z=\left\{
  \begin{array}{ll}
    0  & |x| > \triangle h  \\
    B_0 \cos{\frac{x \pi}{2 \triangle h }}  & |x| < \triangle h \,\,
    ,
  \end{array}
  \right.
\end{equation}

\noindent where $B_0$ and $\triangle h$ are the background
magnetic field and the half width of current sheet,
respectively. In order to make our simulation more
realistic, we use the temperature distribution of the quiet-Sun
atmospheric model VALC as shown in Figure~\ref{fig_ini} (left
panel). Using the hydrostatic equilibrium equation and the modified
Saha and Boltzmann formula for calculating the number density of
electrons iteratively, we get the initial number density of hydrogen
($n_H$) and electrons ($n_e$) as shown in the middle and the right
panels of Figure \ref{fig_ini}, respectively. From these figures we
can see that our initial model is very close to the VALC model. In
order to initiate the magnetic reconnection, we assume that an
anomalous resistivity \citep{Cramer1979}, with the form $\eta
=\eta_0 \cos [ x \pi / 2 \triangle h] \cos [ (y - h_r)\pi /2
\triangle h]$, is localized in a small region $|x| \leq \triangle
h$, $|y - h_{r}| \leq \triangle h$ (the small rectangle in Figures
\ref{mag_field}, \ref{fig_sho} and \ref{fig_Case_jz}), where $h_{r}$
stands for the height of the reconnection point (X-point) in the
$y$-axis, $\triangle h$ is the half width of the resistivity region
and $\eta_0$ is the resistive amplitude. It is noted that the half
width of the resistivity region is taken to be the same as the half
width of the current sheet.

Owing to the symmetry, the calculation is performed only in
the right half region, with the symmetry boundary condition being
set on the left-hand side. The right and the upper boundaries
are treated as open boundaries. Line-tying conditions are
applied to the bottom boundary. It is noted that $\rho$ should
satisfy the hydrostatic equilibrium ($\partial p / \partial y  +
\rho g = 0$) to reach a balance between the gravity and the pressure
gradient force at the upper and the lower boundaries. The
calculation domain is discretized into $800 \times 400$ grid meshes.
The grid sizes are $\triangle x = 1.25$ km and $\triangle y = 5$ km,
which are much less than the pressure scale height ($> 100$ km).
With gravity, ionization and radiation being considered,
and with heat conduction being neglected, the numerical simulations are
performed with the CIP-MOCCT scheme.

\section{NUMERICAL RESULTS}
\label{NUMERICAL RESULTS}
\subsection{Dynamic Process}
\label{Dynamic Process}

\begin{figure}
\begin{center}
\includegraphics[width=210pt]{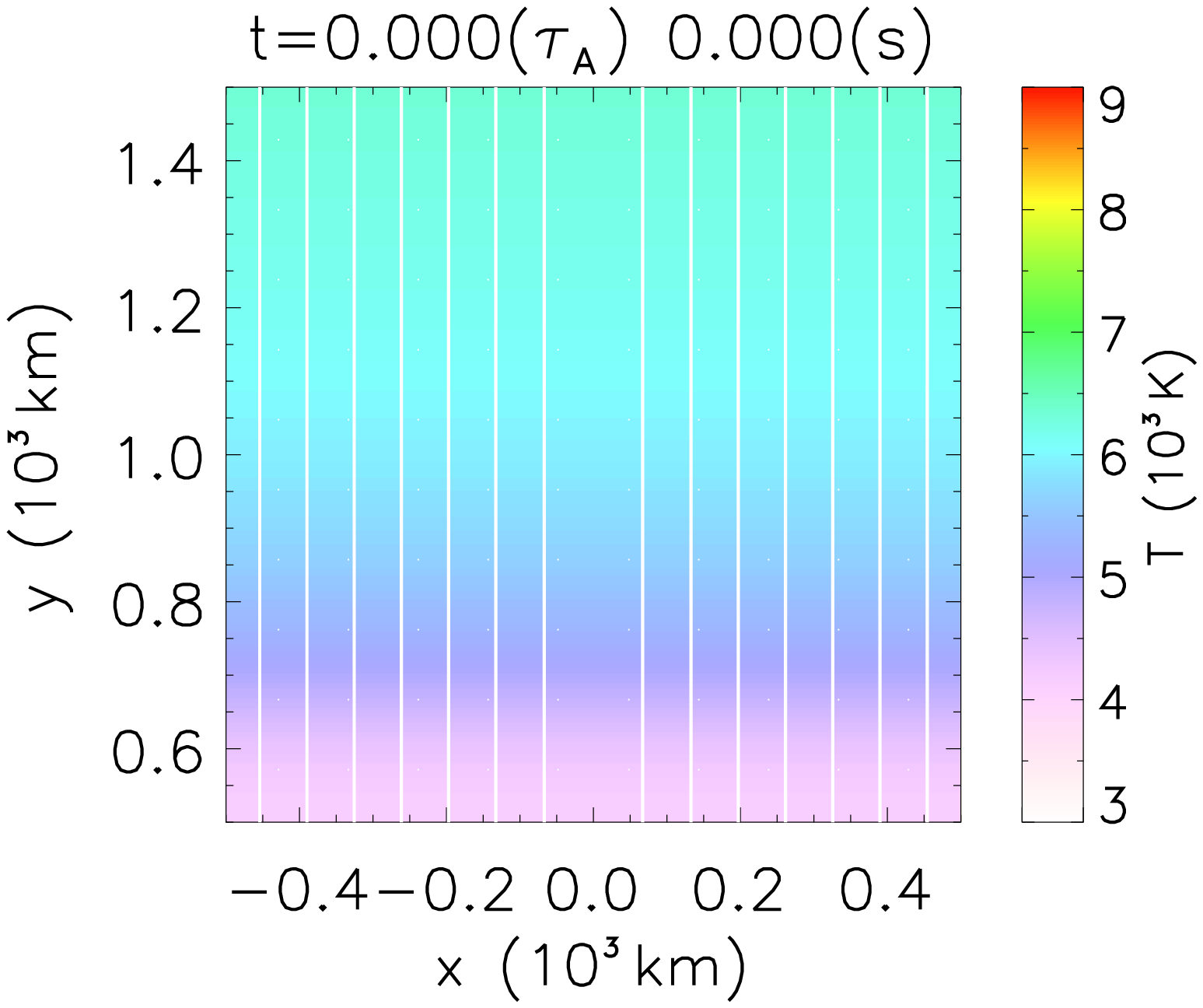}
\includegraphics[width=210pt]{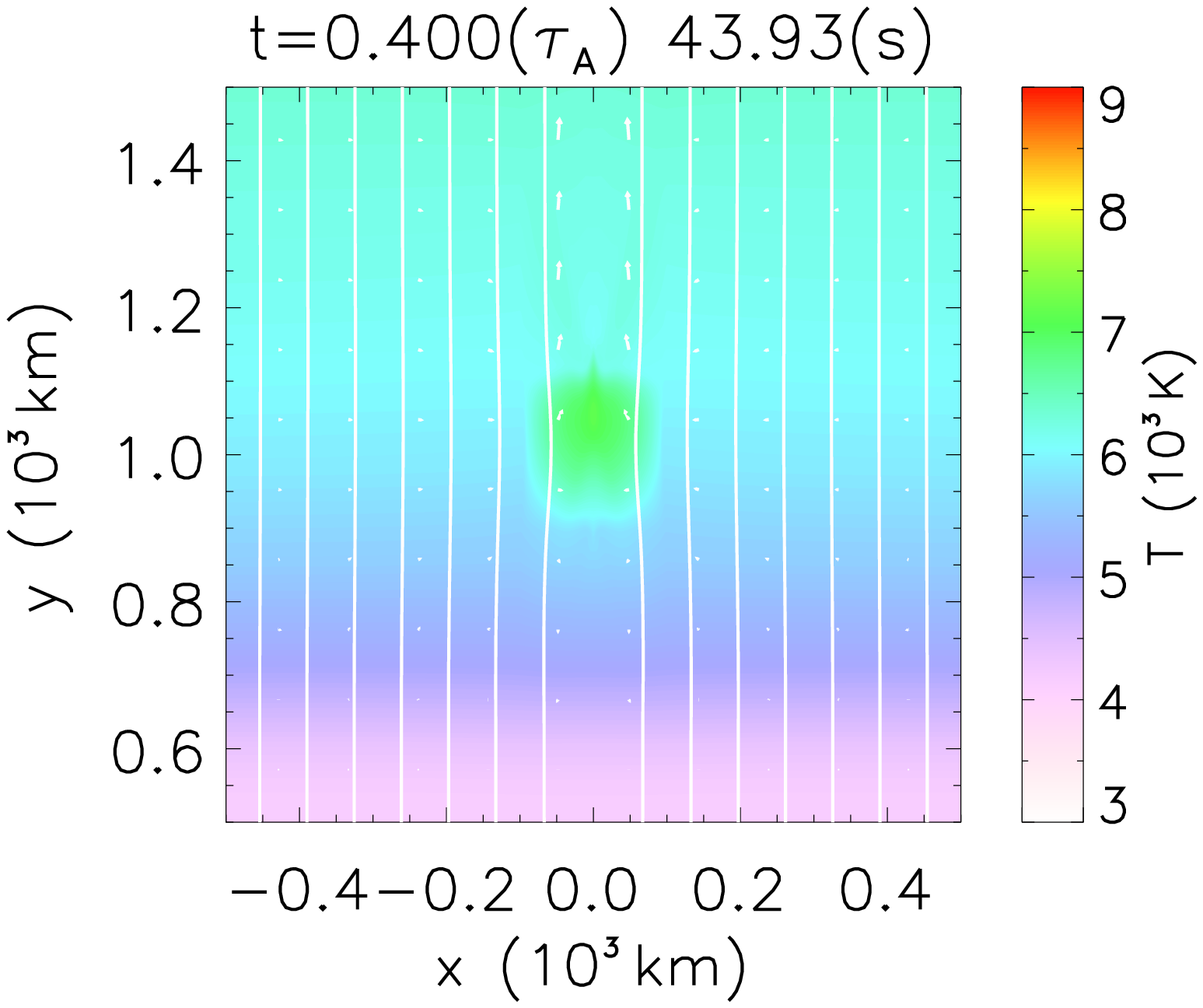}
\includegraphics[width=210pt]{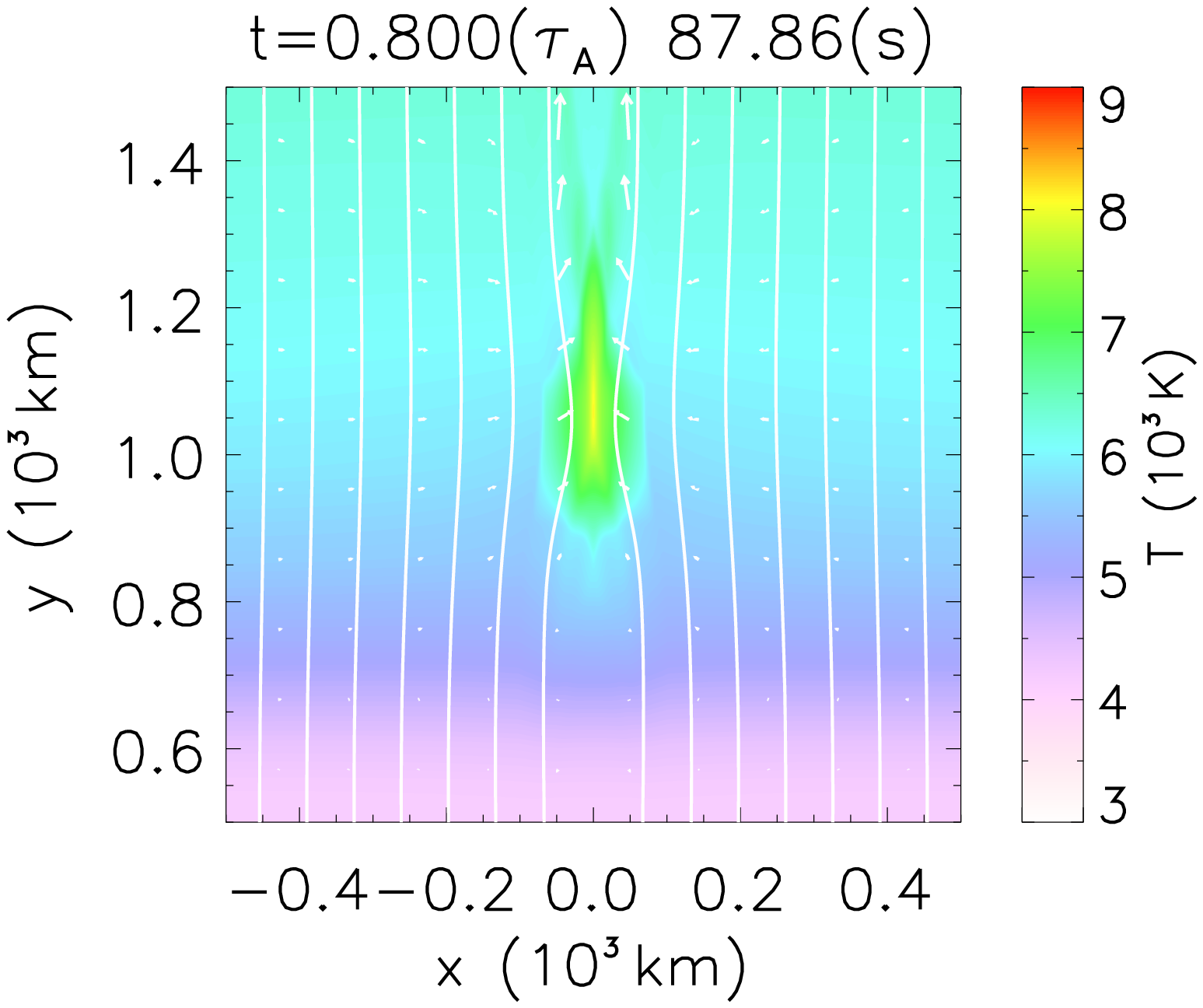}
\includegraphics[width=210pt]{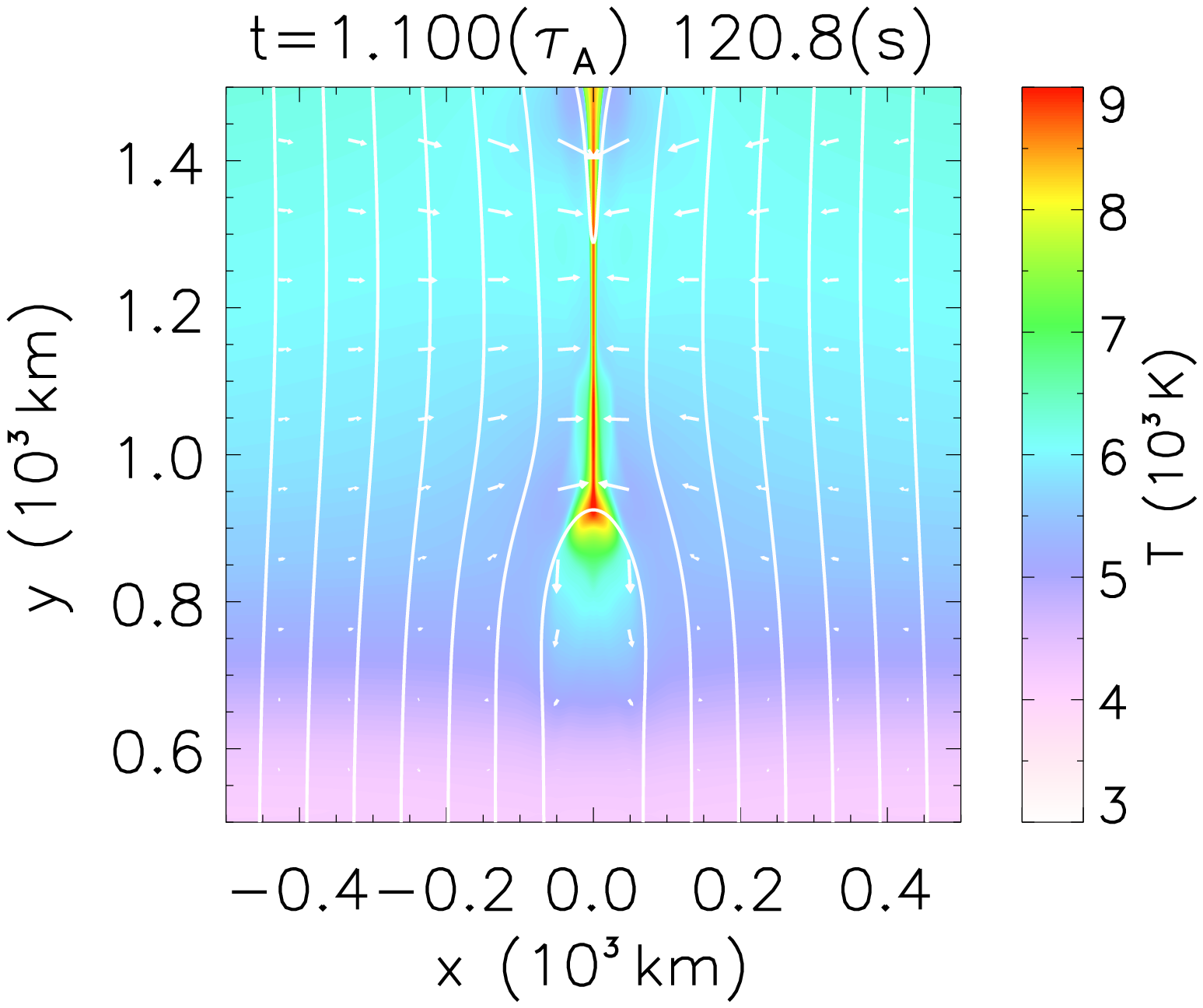}
\end{center}
 \caption{Temperature distributions (color scale), projected
  magnetic field (solid lines) and velocity field (vector arrows) at
  $t$ = 0  and $t$ = 0.4 (upper panels) and $t$ = 0.8 and $t$ = 1.1
  $\tau_{A}$ (lower panels).} \label{fig_res}
\end{figure}

Generally speaking, the height of the X-point may change as the
reconnection proceeds \citep{Takeuchi2001, von_Rekowski2008}. In our
simulation, we found that the X-point could move up or down
by $10-40$ km, but still within the resistivity region. For this
reason, we do not need to change the position of the
resistivity region. In order to illustrate the dynamic process, this
subsection describes the simulation results in a typical case, where
the background magnetic field $B_{0}$ is taken to be 50 G, the
anomalous resistivity $\eta_{0}$ be 17.7 $\Omega$ m, the X-point
height $h_{r}$ be 1000 km, and the half width of the resistivity
region $\triangle h$ be 100 km. Figure~\ref{fig_res} depicts the
results of the simulation at different times. In this figure the color
stands for the temperature, solid lines for magnetic field and
arrows for velocity.

\begin{figure}
\begin{center}
\includegraphics[width=210pt]{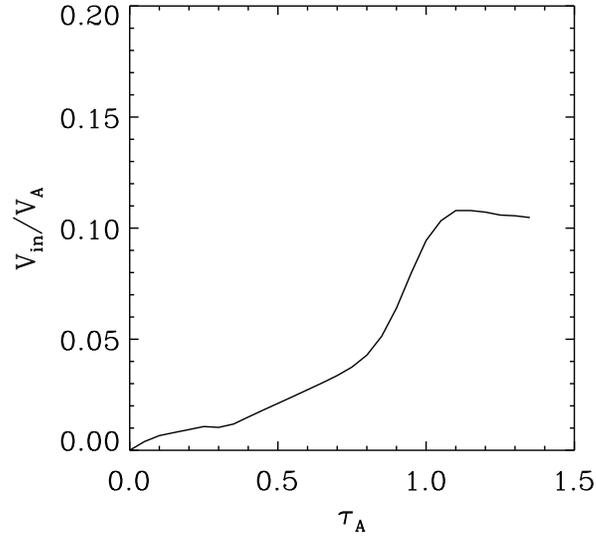}
\end{center}
 \caption{Evolution of the magnetic reconnection rate $|V_{in}| /
 V_A$. The rate reaches the maximum at the time of 1.1 $\tau_{A}$
 (120.8 s).}
 \label{fig_rec}
\end{figure}

As the anomalous resistivity sets in, we have a fast reconnection
process similar to that predicted by \cite{Petshek1964}. Figure
\ref{fig_rec} shows the evolution of the reconnection rate, which is
defined as $|v_{in}| / v_A$, where $v_{in}$ is the maximum inflow
velocity along the $x$-direction at the height of the X-point. It is
found that the simulation stops at $t= 1.35\tau_{A}$ (148.2
s), probably due to some numerical instability. The reconnection
rate reaches the maximum, 0.11, at $t=1.1\tau_{A}$ (120.8 s). After
that, the reconnection rate remains almost constant around
0.1. The lifetime of the reconnection simulated in this case (148.2
s) is shorter than the typical lifetime of microflares, i.e.,
$10-30$ minutes. Maybe our results can only represent the
pre-flare phase and the rise phase.

\begin{figure}
\begin{center}
\includegraphics[width=300pt,height=300pt]{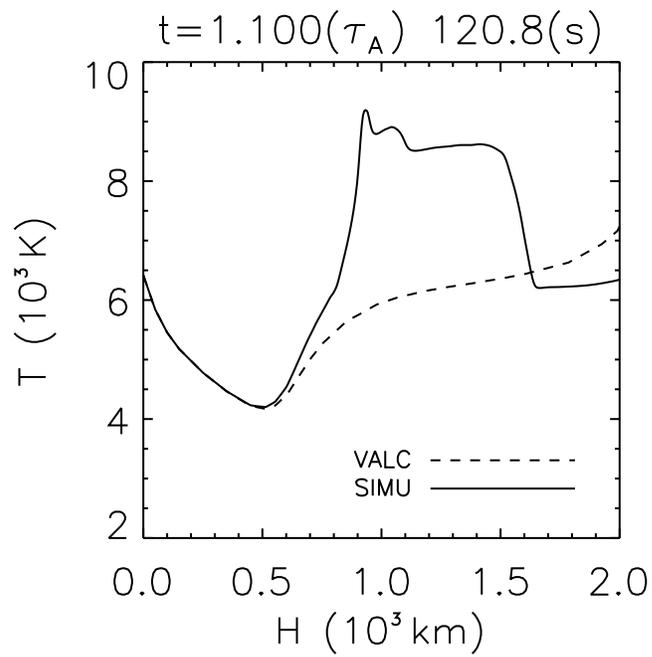}
\end{center}
\caption{One-dimensional $T$ plot (solid line) along the y-axis at
 the time of the maximum reconnection rate (1.1 $\tau_{A}$). The
 initial temperature plot (dashed line), i.e. VALC temperature
 distribution, is also shown.}
  \label{fig_tem}
\end{figure}

\begin{figure}
\begin{center}
\includegraphics[width=210pt]{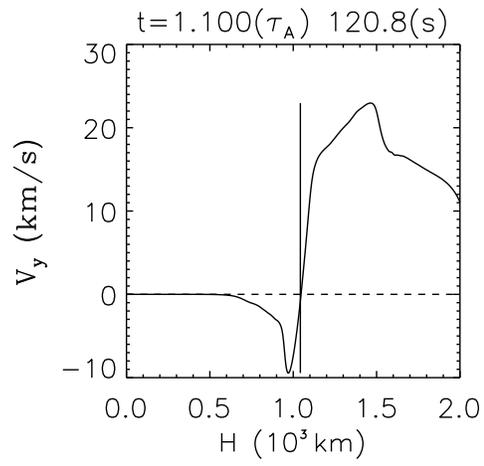}
\end{center}
 \caption{One-dimensional $V_y$ plot (solid line) along the y-axis at
  the time of maximum reconnection rate (1.1 $\tau_{A}$). The height
  of the X-point is marked by a vertical line. The initial $V_y$ plot
  (dashed line) is also shown.} \label{fig_vy}
\end{figure}

During the reconnection, the temperature enhancement in the
resistivity region is about $2000-3000$ K as shown in Figure
\ref{fig_tem}. Accelerated by the magnetic tension force, the hot
plasma together with the frozen-in reconnected field lines are
ejected from the resistivity region to form upward and downward
reconnection jets. The maximum velocity of the upward and downward
jets are about $V_{up} = 23.0$ km s$^{-1}$ and $V_{down} = 9.4$ km
s$^{-1}$, respectively as shown in Figure~\ref{fig_vy}. Because of
the sharp density gradient in the lower atmosphere, the speed of the
upward jet is faster than that of the downward one. Compared with the
local Alfv\'en speed ($V_A = 18.2$ km s$^{-1}$) at the height of 1000
km, we find that the velocity of the outflow is approximately
the local Alfv\'en speed in the inflow region. The downward jet can only reach the height of 600 km, after which it is slowed down significantly. Furthermore, the reconnection X-point also rises by about 40 km during the reconnection process.

The hot plasma jet ejected along the $y$-axis is mainly heated by
the Joule dissipation in the resistivity region. However, we can not
ignore the heating by the slow-mode shocks. In our simulation, the
heating occurs not only in the resistivity region but also outside
the region as shown in Figure~\ref{fig_sho}, which shows the current
density ($z$-component) distribution in the left panel and the
temperature distribution in the right panel. It can be seen that a
pair of slow-mode shocks, which are characterized by the high
current density, are formed at the interface between the
reconnection inflow and outflow, across which the plasma temperature
increases drastically. It indicates that besides the Joule
dissipation, the slow-mode shock may be another effective
way to heat plasma.

\subsection{Parameter Dependence}
\label{Parameter Dependence}

In our simulation, the most important parameters are the background
magnetic field, the anomalous resistivity, and the height of the
X-point (i.e., the reconnection site). In order to show the
influence of different parameters on the results, we performed
extensive simulations by changing one parameter at a time with
others being fixed.

\begin{figure}
\begin{center}
\includegraphics[width=200pt,height=200pt]{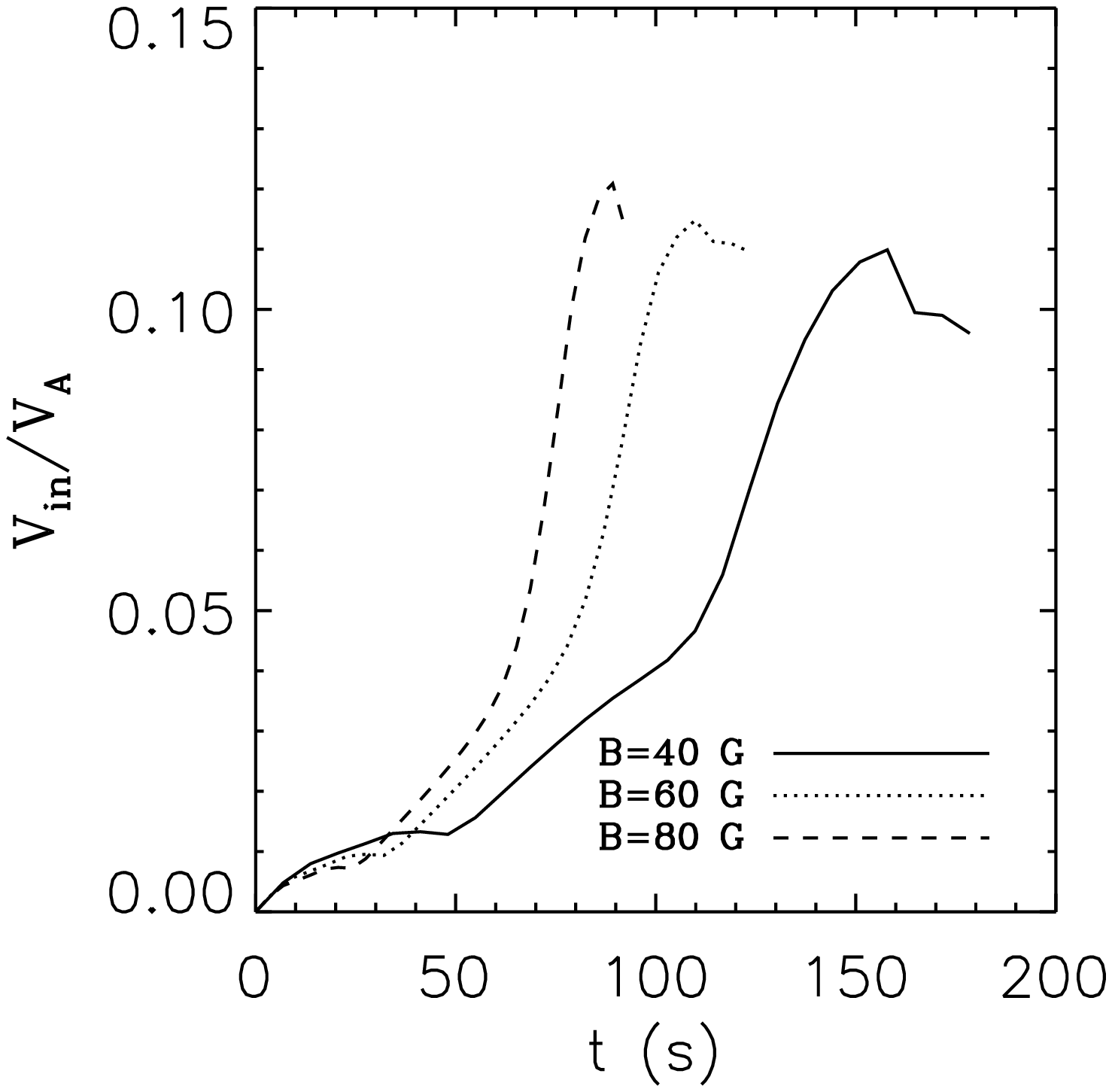}
\includegraphics[width=200pt,height=200pt]{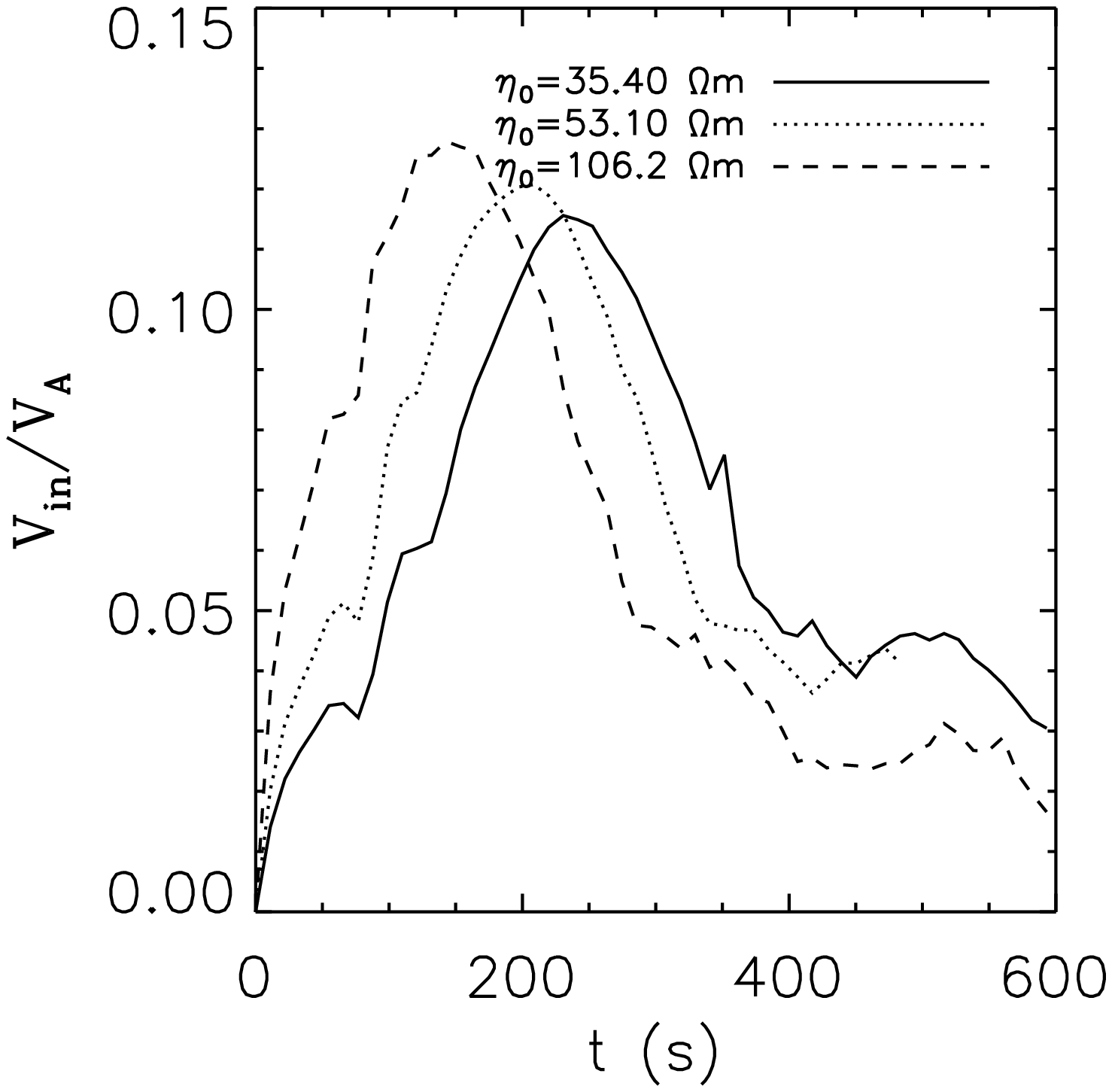}
\end{center}
\caption{Evolution of the magnetic reconnection rate for different
 initial background magnetic fields ($B$) and anomalous resistivity
 ($\eta_0$). The height of the initial X-point is 1000 km. }
  \label{fig_Par_Rate}
\end{figure}

Figure \ref{fig_Par_Rate} shows the evolution of the magnetic
reconnection rate for different initial background magnetic fields
and anomalous resistivities, with the height of the initial X-point
being 1000 km. It can be seen that the magnetic
reconnection rate can reach the maximum much faster when we set a
stronger initial magnetic field. It is also found that the large
anomalous resistivity can shorten the evolution time from the
initial stage to the time when the reconnection rate reaches the
maximum. It should be noted here that, although the resistivity in the chromosphere is $\sim3$ orders of magnitude larger than that in
the corona, it is still too small from the simulation point of view. A
so small resistivity would lead to a very thin diffusion layer near the reconnection X-point, which could not be resolved by the current
numerical mesh. Therefore, the values of the anomalous
resistivity we used are several orders of magnitude larger than the
classical Spitzer resistivity (e.g., Kovitya \& Cram 1983).

\begin{figure}
\begin{center}
\includegraphics[width=200pt,height=200pt]{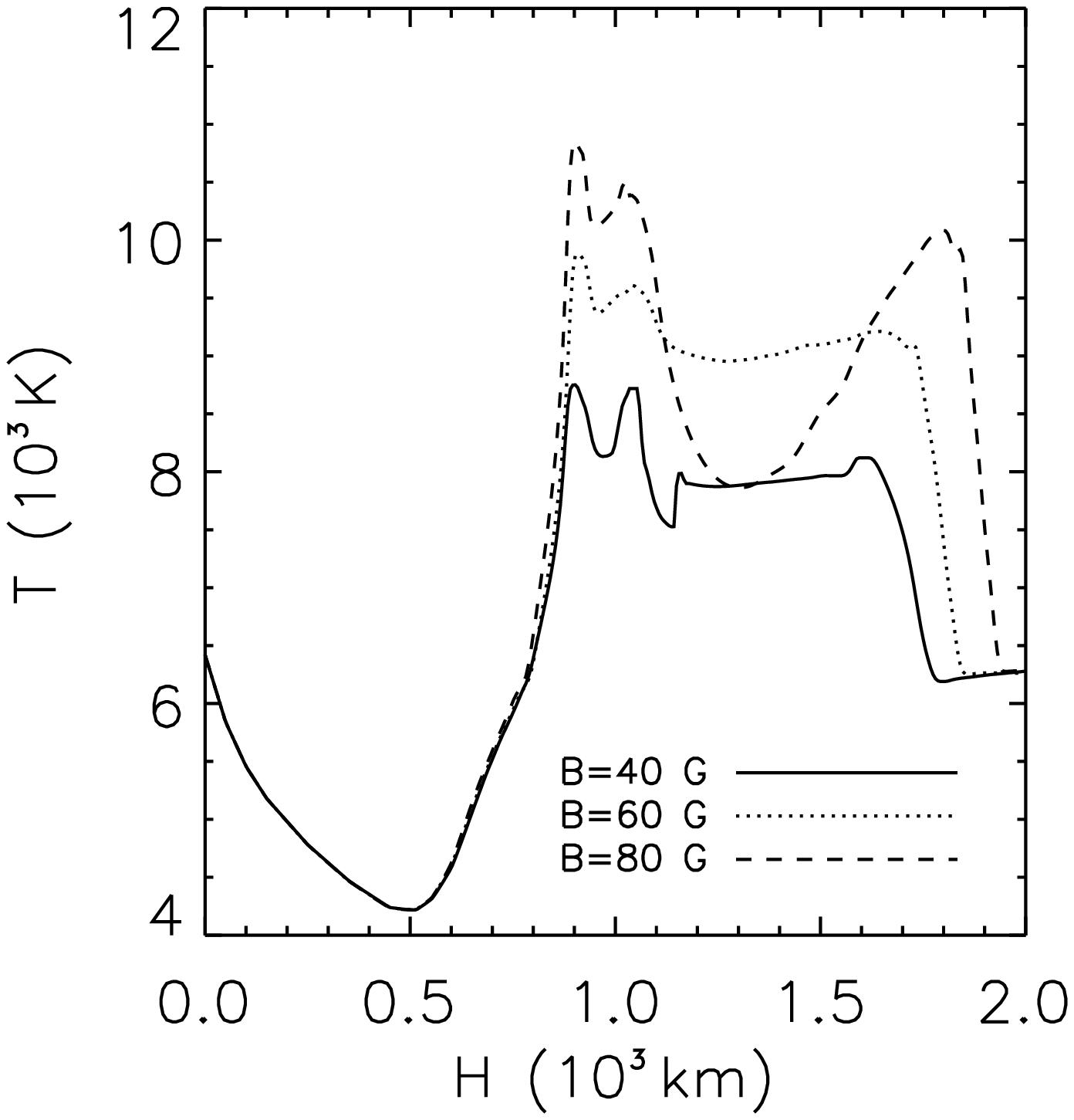}
\includegraphics[width=200pt,height=200pt]{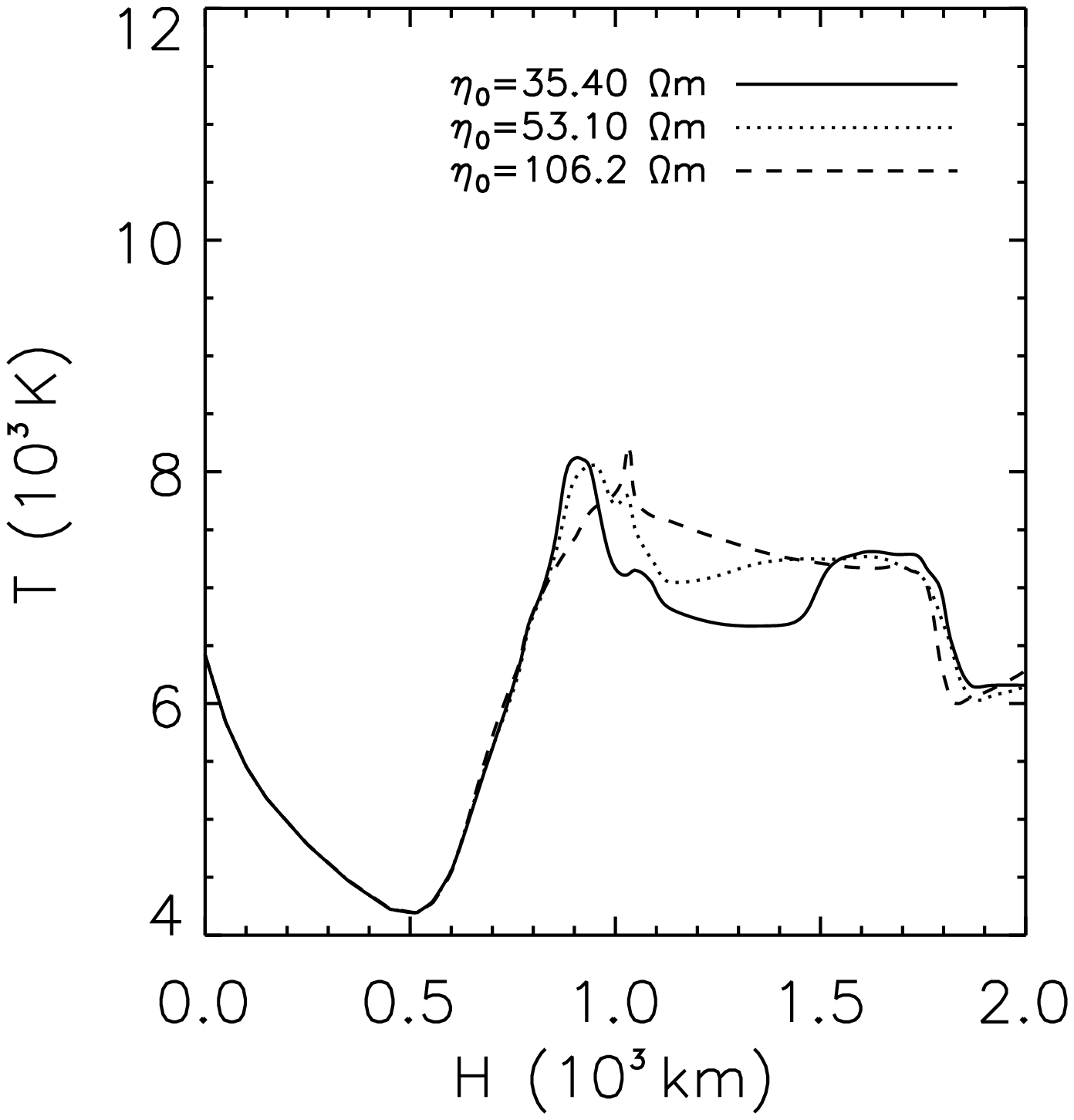}
\end{center}
\caption{Temperature distributions for different initial background
 magnetic field ($B$) and anomalous resistivity ($\eta_0$) at the
 time of maximum rate of magnetic reconnection which have been shown
 in Figure \ref{fig_Par_Rate}. The height of the initial X-point
 is 1000 km.}
  \label{fig_Par_Temp}
\end{figure}

Figure \ref{fig_Par_Temp} shows the temperature distributions at the
peak time of the magnetic reconnection rate. Our results indicate
that the temperature enhancement is sensitive to the magnetic field.
However, in all cases our results are consistent qualitatively with the observed temperature enhancement in chromospheric microflares \citep{Fang2006a}. A quantitative comparison between the simulation and the observation will be described in Section 3.4.

\subsection{Scaling law}
\label{Scaling law}

\begin{figure}
\begin{center}
\includegraphics[width=150pt,height=150pt]{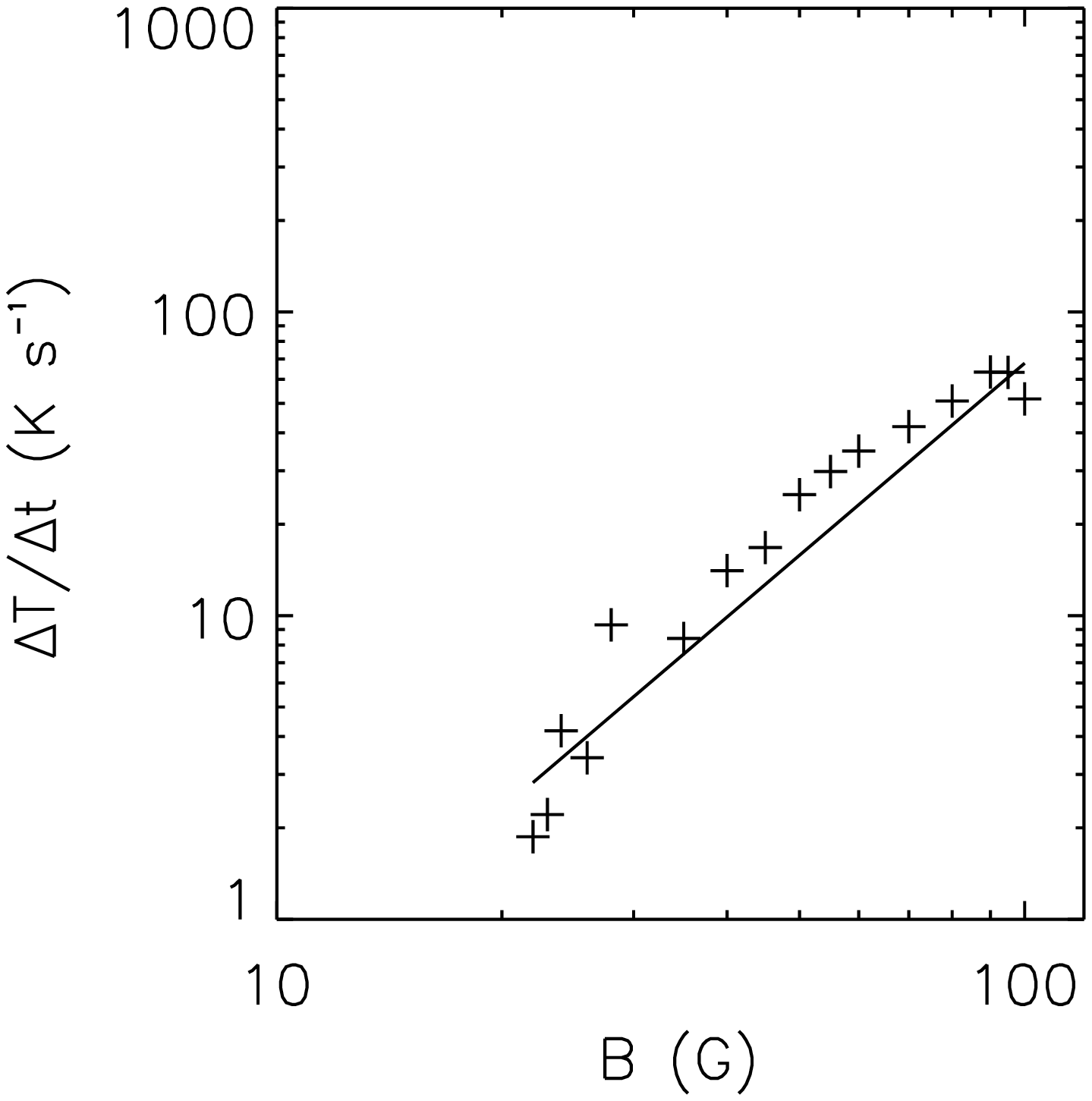}
\includegraphics[width=150pt,height=150pt]{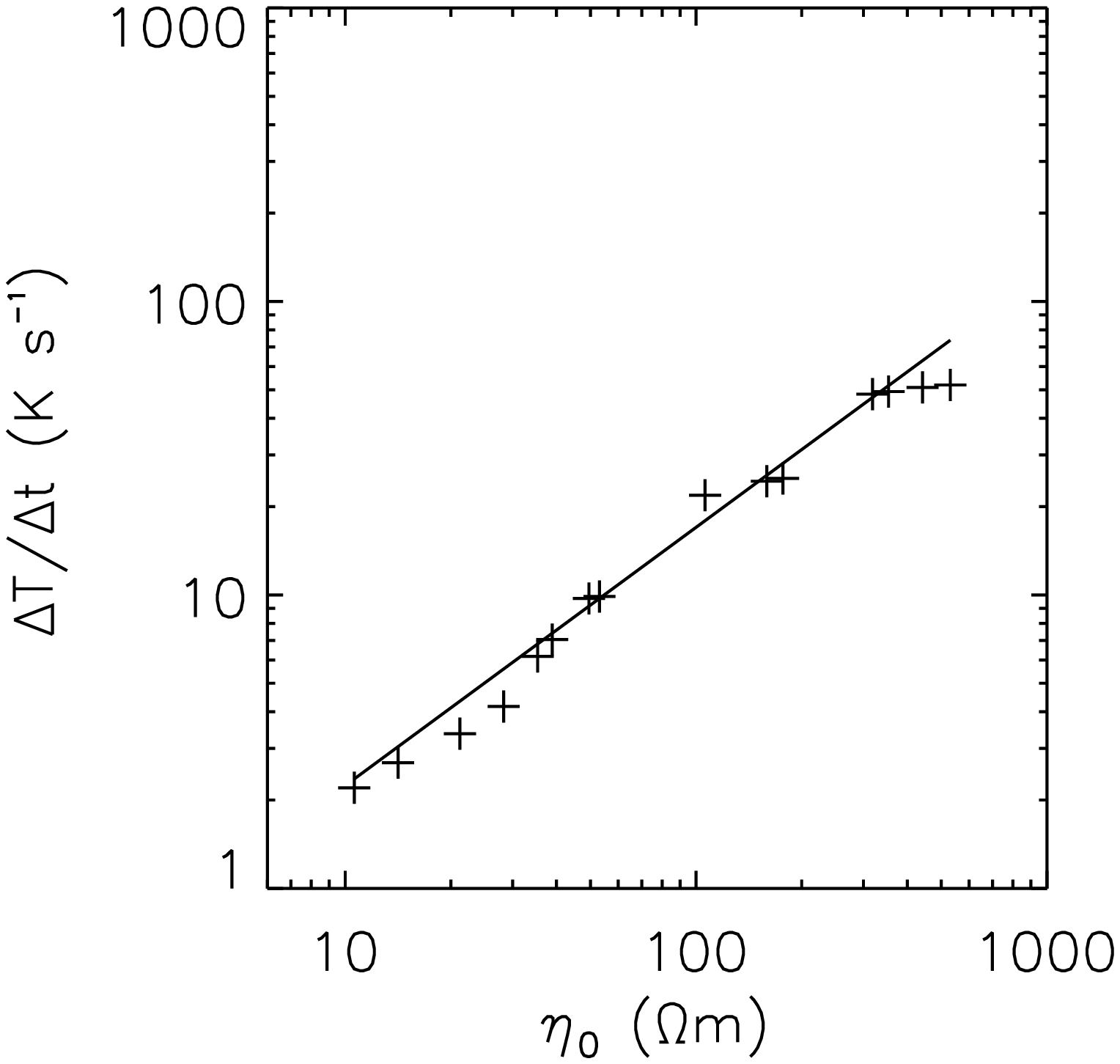}
\includegraphics[width=150pt,height=150pt]{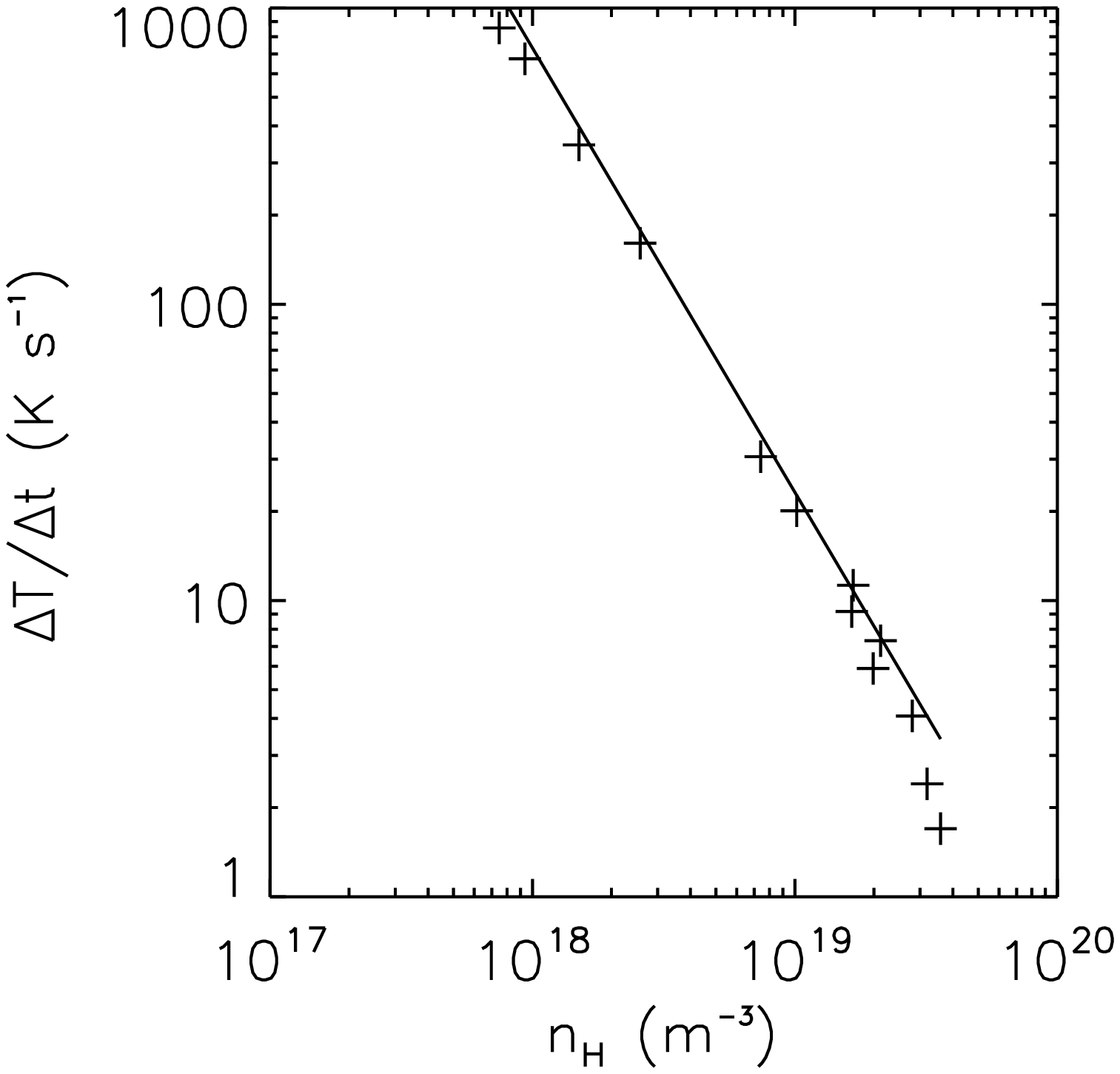}
\end{center}
\caption{Temperature enhancement per unit time as a function of
 the background magnetic field $B$ (left panel), the anomalous
 resistivity $\eta_0$ (middle panel), and the number density of
 hydrogen atom $n_H$ (right panel). In the left panel, the initial
 reconnection height is 1000 km and the anomalous resistivity is
 17.7 $\Omega$m. In the middle panel, the initial reconnection
 height is 1000 km and the magnetic field is 25 G. In the right
 panel, the anomalous resistivity is 17.7 $\Omega$m and the
 magnetic field is 25 G. }
  \label{fig_sca}
\end{figure}

The ranges of the variables such as the background magnetic
field ($B_0$), anomalous resistivity ($\eta_0$) and the number density
of hydrogen ($n_H$) are very wide in observations. As a result, the
maximum temperature enhancement ($\Delta T$) and the reconnection
lifetime ($\Delta t$) may change greatly. In order to find
a scaling law to relate $\Delta T$ and $\Delta t$ to the
above-mentioned 3 variables, we change the variables one by one and
perform a series simulations. As illustrated by Figure
\ref{fig_sca}, the dataset can be well fitted by the following
formula:

\begin{equation}
\frac{\triangle T}{\triangle t} \approx 3.7 (\frac{n_H}{3.4 \times
 10^{19} \ {\rm m}^{-3}}) ^{-1.5} (\frac{B_0}{25 \ {\rm G}})^{2.1}
 (\frac{\eta_0}{17.7 \ {\rm \Omega ~m}})^{0.88} \ {\rm K~s}^{-1} \,\, .
 \label{sca_1}
\end{equation}

From the energy point of view, as the reconnection proceeds, the Joule
dissipation is consumed partly by the plasma radiation and ionization,
with the rest being converted into the plasma heating. Therefore, we
have

\begin{equation}
n_H k \triangle T = \frac{J^2}{\sigma}\triangle t - n_H \alpha(y)
n_e f(T)\triangle t + n_H \alpha(y) [n_e f(T)]_{t=0}\triangle
t-[n_e-(n_e)_{t=0}] \chi_H \,\, ,
 \label{sca_2}
\end{equation}

\noindent
where the expressions of $n_e$, $\alpha(y)$, $f(T)$ are
given by Equations (\ref{Ne-1}), (\ref{Rad-2}), and (\ref{Rad-3}).
The items on the right-hand side of the Equation (\ref{sca_2}) are
the Joule dissipation, the radiative losses, the pre-heating and the
ionization energy, respectively. If we assume that the plasma
radiation and ionization are negligible, Equation (\ref{sca_2}) can
be rewritten in a simple form as

\begin{equation}
 n_H k \triangle T \approx \frac{J^2}{\sigma} \triangle t = \eta
 \frac{(\nabla \times B_0)^2}{{\mu_0}^2}\triangle t \sim \eta
 \frac{B_0^2}{{\mu_0}^2 L^2} \triangle t \,\, ,
 \label{sca_3}
\end{equation}

\noindent where $\mu_0$ is the vacuum permeability, $L$ is the scale
of the reconnection region. Since in our simulation $L$ is fixed as a
constant, we can get a simple theoretical formula as follow:

\begin{equation}
\frac{\triangle T}{\triangle t} \propto \frac{B_0^2 \eta}{n_H}\,\, .
\label{sca_4}
\end{equation}

It can be seen that the formula (\ref{sca_1}) given by our simulations
is similar to the formula (\ref{sca_4}) except the relatively bigger
deviation of the power index of $n_H$. The deviation is probably because
the radiative loss and ionization items, which are related
to $n_H$ but not to $B_0$ and $\eta_0$, have been considered in our
simulations but are neglected in the analytical form. A further study
will be done in order to find a more realistic scaling law in the future.

\subsection{Comparison with a Bright Microflare Event}
\label{Compare}

\begin{figure}
\begin{center}
\includegraphics[width=150pt,height=150pt]{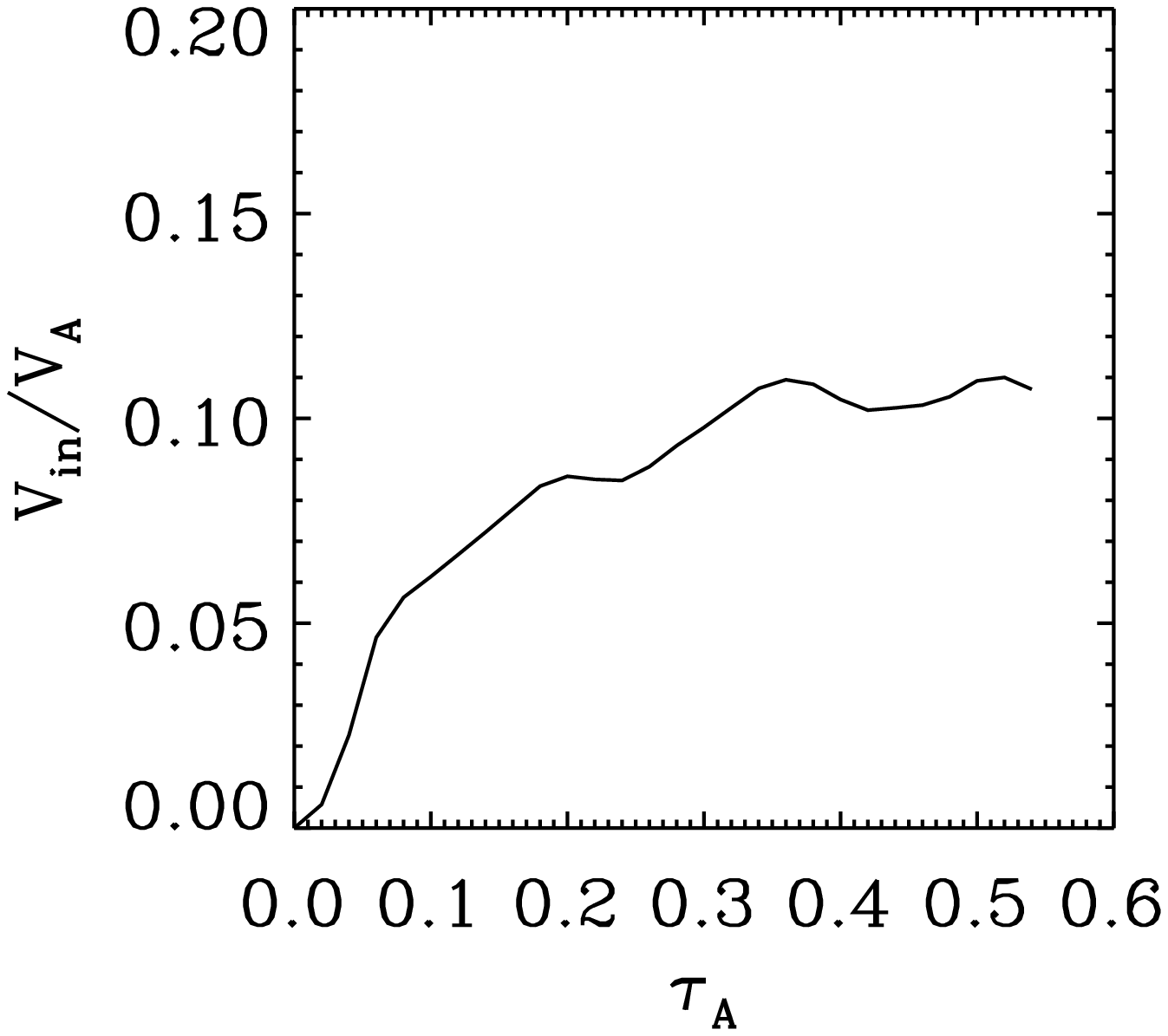}
\includegraphics[width=150pt,height=150pt]{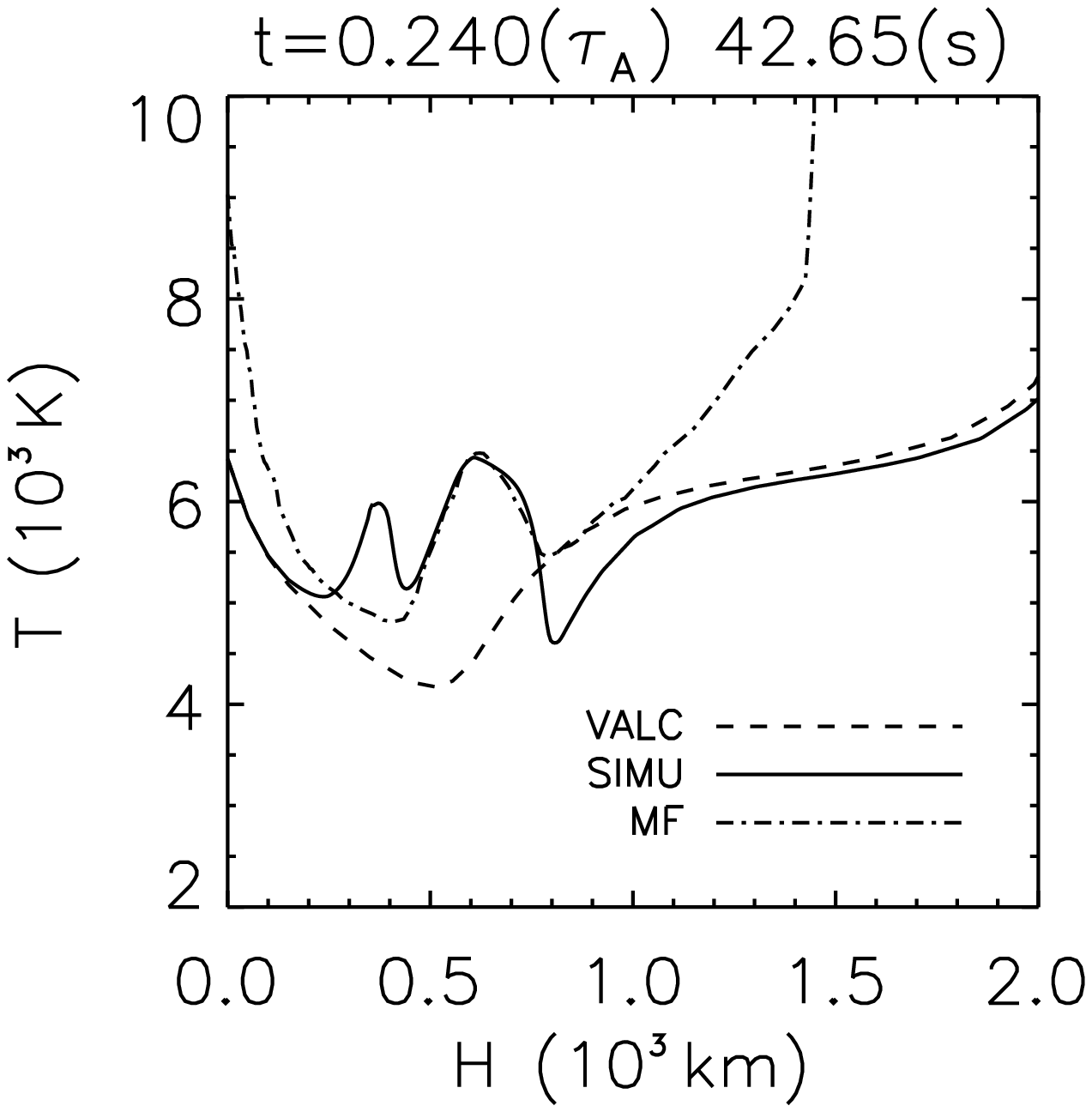}
\includegraphics[width=150pt,height=150pt]{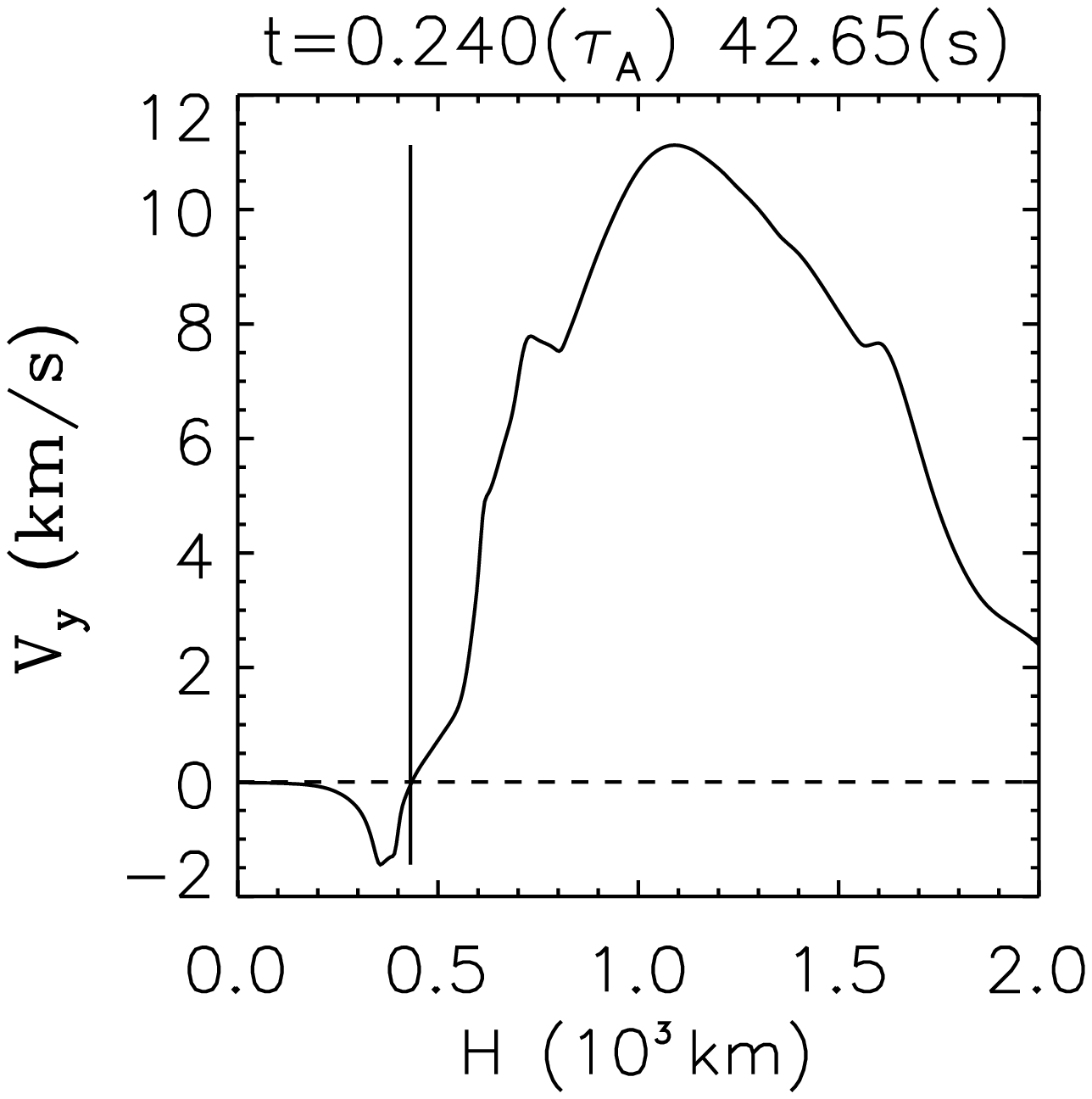}
\end{center}
\caption{Magnetic reconnection rate (left panel), $T$ (middle panel)
 and $V_y$ (right panel) plot (solid line) at the time 0.24 $\tau_A$.
 The initial conditions of $T$ and $V_y$ are also shown (dashed lines).
 The height of X-point is marked by the vertical line in the right
 panel and the temperature distribution of bright microflare (MF)
 is given by dash-dotted line in the middle panel.}
  \label{fig_Case}
\end{figure}

Fang et al. (2006a) analyzed five well-observed microflares and
computed the thermal semi-empirical models for two typical
microflares, one was bright and the other was
faint. In oder to get the semi-empirical models, they first assume the
temperature and turbulent velocity distributions, and then derive the
density distribution. Based on these distributions they can calculate
the H$\alpha$ and \ion{Ca}{2} 8542 {\AA} line profiles. The atmospheric
model is adjusted until the calculated line profiles to match the
observed ones. The temperature distribution of the
bright microflares model is shown in the middle panel of
Figure~\ref{fig_Case}. In order to reproduce the temperature
enhancement in the model, we set the initial background magnetic
filed to be 300 G, the anomalous resistivity to be 159.3 $\Omega$ m,
and the initial reconnection height to be 500 km. Since the
magnetic field is strong, the total evolution time is short. The
evolution of the reconnection rate, the temperature enhancement, and
the velocity along the $y$-axis ($v_y$) are shown in
Figure~\ref{fig_Case}.

After $t= 0.2\tau_A$, the reconnection rate reaches about 0.1. We
choose the temperature distribution at 0.24 $\tau_A$ to compare with
the semi-empirical model. It is shown that there are two
peaks in our simulation results. The smaller one is located at the
height of 350 km and the bigger one is almost the same as the
semi-empirical model and located at the height of 600 km. The
possible reason for this phenomenon is slow-mode MHD shock heating
as shown in Figure \ref{fig_Case_jz}. Below 800 km, the temperature
profile of our simulation is close to that of the
semi-empirical model. However, above 800 km, the temperature in the
semi-empirical model becomes much higher than that of our
simulation. The discrepancy might result from the fact that
our simulations are performed for a relatively short period
such that the reconnection jet has not yet reached the
corona. Therefore, our result can only give a reasonable explanation
for the temperature bump of the microflare mainly in the rise phase.

The right panel of Figure~\ref{fig_Case} shows the one-dimensional
plot of $v_y$. From this figure it can be seen that the height of
the reconnection X-point, which is characterized by the null
velocity, has dropped from 500 km at $t=0$ to 450 km at
$t=0.24\tau_A$.  The velocity of the upward flow is about 11.1 km
s$^{-1}$, which is much larger than that of the downward flow (1.8
km s$^{-1}$). The downward jets can only reach the
height of about 200 km.

\section{DISCUSSION AND SUMMARY}
\label{Disc}

More and more theoretical works have indicated that magnetic
reconnection in the solar lower atmosphere can produce chromospheric
microflares \citep{Tand1988, Liu2004, Fang2006a}, as well as other
small activities like Ellerman bombs \citep{Fang2006b, Ding1998,
Watanabe2008}. With gravity, ionization and radiation being
considered, we performed 2.5D MHD simulations. Since we
have used the temperature distribution of the quiet-Sun
atmospheric model VALC \citep{Vernazza1981} and considered the
helium abundance ($10\%$), our simulations are realistic to some
extent.

In our simulation, we calculate the magnetic reconnection rate by
using the definition $R = v_{in}/v_A$, rather than the
formula $R = d \psi / dt$ used by \cite{Chen1999, Chen2001} and $R =
|\eta J_z|$ used by \cite{Yokoyama2001}, where $\psi$ means the
magnetic flux function and $J_z$ stands for the $z$-component of
current density. If we use the formula $R = d \psi / dt$ or $R=|\eta
J_z|$, the reconnection rate has a large pulse, which can be
greater than 1.0 at the beginning of the evolution. It is
non-physical, as mentioned by \cite{Yokoyama2001}. However, it
should be noted that all of these methods can give a
similar result after enough time of evolution.

In the subsections~\ref{Dynamic Process} and~\ref{Compare}, we
studied the magnetic reconnection with the X-point at different
heights, e.g., 500 km and 1000 km. Our results, in the case of 500
km, can well reproduce the temperature enhancement in the
semi-empirical model \citep{Fang2006a}. At the heights of 500 km and
1000 km , the temperature enhancement is 1000-2000 K and 2000-3000
K, respectively. In the case of 1000 km, we found that the
height of X-point can move up by tens of kilometers.
This is different from the results of the case of 500 km,
where the X-point moves down. It is noted that the temperature
enhancement we got in our simulations is several thousand Kelvin,
and our results can only represent the rise phase of chromospheric
microflares since our simulations were performed only until the upward
reconnection jet reaches the bottom of the transition region. Of course, those microflares with strong EUV, X-ray, and microwave
emissions may occur at the lower corona. In this case, the emissions in X-rays, EUV and microwave can be naturally explained.

It is worth noting that our simulations have some limitations. Even
though we limited the computational height below $2000$ km,
the plasma $\beta$ at the bottom boundary is still six orders of
magnitude larger than that at the top boundary. It causes
some numerical instability in our simulations. Besides, our initial
magnetic configuration is relatively simple. Moreover, as
it is well known, the current sheet should be very thin in the
real situation, e.g., less than hundreds of meters, while in our
simulation the minimum grid sizes are $\triangle x = 1.25$ km and
$\triangle y = 5$ km, which are too large to simulate the current
sheet. All these contribute to the fact that our
simulations can not reproduce the real observations in details.
Further improvement of this work is expected.

In summary, we give the conclusions as follows:

1. Our 2.5D MHD simulations can reproduce the temperature
enhancement in chromospheric microflares qualitatively. The
temperature increase in the cases when the reconnection point is at
the height of 500 km and 1000 km can reach 1000-2000 K and 2000-3000
K, respectively.

2. The free parameters in our 2.5D simulation are the
background magnetic field ($B_0$) and the anomalous resistivity
($\eta_0$). We have found that the temperature enhancement is
sensitive to the background magnetic field, while the total evolution
time is sensitive to the magnitude of anomalous resistivity.

3. Our simulation results indicate that the velocity of the upward
jet is much larger than that of the downward jet, and the X-point may
exhibit downward or upward motions.

4. We have performed a parameter survey, and found that the
temperature enhancement and the total evolution time in the
chromospheric reconnection follow the scaling law:

\begin{equation}
\frac{\triangle T}{\triangle t} \approx 3.7 (\frac{n_H}{3.4 \times
 10^{19} \ {\rm m}^{-3}}) ^{-1.5} (\frac{B_0}{25 \ {\rm G}})^{2.1}
 (\frac{\eta_0}{17.7 \ {\rm \Omega~m}})^{0.88} \ {\rm K~s}^{-1} \,\, .
\end{equation}

\acknowledgments The computations were done by using the HP ProLiant
BL260C G5 Blade system at the Department of Astronomy of Nanjing
University of China. This work is supported by the National Natural
Science Foundation of China (NSFC) under the grant numbers 10221001,
10878002, 10403003, 10620150099, 10610099, 10933003 and 10673004, as
well as the grant from the 973 project 2006CB806302.

\clearpage

\begin{table}[t]
\begin{center}
 \caption{Normalization Units}
\label{Normalization Units}
\begin{tabular}{cccc}
\hline
\hline
Variable & Quantity & Unit & Value \\
\hline
$x,y$        & Length         & $L_0$                                  & $2000$ km                          \\
$T$          & Temperature    & $T_0$                                  & $6000$ K                           \\
$\rho$       & Density        & $\rho_0$                               & $6.0 \times 10 ^ {-8}$ kg m$^{-3}$ \\
$p$          & Pressure       & $p_0=\rho_0 T_0 \kappa_B / m_H$        & $2.97$ N m$^{-2}$                  \\
$\mathbf{V}$ & Velocity       & $V_0=(T_0 \kappa_B / m_H)^{1/2}$       & $7.035$  km s$^{-1}$               \\
$\mathbf{B}$ & Magnetic field & $B_0=(\mu_0 \rho_0 T_0 \kappa_B / m_H)^{1/2}$ & $19.3$ G                    \\
$t$          & Time           & $t_0=L_0 (T_0 \kappa_B / m_H)^{-1/2}$         & $284 $ s                    \\
\hline
\end{tabular}
\end{center}
\end{table}

\clearpage

\clearpage

\clearpage

\clearpage

\clearpage

\clearpage

\clearpage

\clearpage

\clearpage

\clearpage

\clearpage

\clearpage

\clearpage


\begin{thebibliography}{}


\bibitem[Brosius \& Holman(2009)]{Brosius2009} Brosius, J.~W.,
  \& Holman, G.~D.\ 2009, \apj, 692, 492

\bibitem[Brown(1973)]{Brown1973} Brown, J.~C.\ 1973, \solphys, 31, 143

\bibitem[Chen et al.(1999)]{Chen1999} Chen, P.~F., Fang, C.,
 Tang, Y.~H., \& Ding, M.~D.\ 1999, \apj, 513, 516

\bibitem[Chen et al.(2001)]{Chen2001} Chen, P.-F., Fang, C.,
  \& Ding, M.-D.~D.\ 2001, ChJAA, 1, 176

\bibitem[Cramer \& Donnelly(1979)]{Cramer1979} Cramer, N.~F.
  \& Donnelly, I.~J.\ 1979, PASAu, 3, 367

\bibitem[Ding et al.(1998)]{Ding1998} Ding, M.~D., Henoux,
  J.-C., \& Fang, C.\ 1998, \aap, 332, 761

\bibitem[Emslie \& Noyes(1978)]{Emslie1978} Emslie, A.~G.,
  \& Noyes, R.~W.\ 1978, \solphys, 57, 373

\bibitem[Fang et al.(2006a)]{Fang2006a} Fang, C.,
 Tang, Y.-H., \& Xu, Z.\ 2006a, ChJAA, 6, 597

\bibitem[Fang et al.(2006b)]{Fang2006b} Fang, C.,
 Tang, Y. H., Ding, M. D., \& Chen, P. F. 2006b, \apj, 643, 1325

\bibitem[Gan \& Fang(1990)]{Gan1990} Gan, W.~Q.,
  \& Fang, C.\ 1990, \apj , 358, 328

\bibitem[Gary \& Zirin(1988)]{Gary1988} Gary, D.~E.,
  \& Zirin, H.\ 1988, \apj, 329, 991

\bibitem[Golub et al.(1974)]{Golub1974} Golub, L.,
  Krieger, A.~S., Silk, J.~K., Timothy, A.~F., \& Vaiana, G.~S.\
  1974, \apjl, 189, L93

\bibitem[Golub et al.(1977)]{Golub1977} Golub, L., Krieger,
  A.~S., Harvey, J.~W., \& Vaiana, G.~S.\ 1977, \solphys, 53, 111

\bibitem[Kovitya \& Cram(1983)]{Kovitya1983} Kovitya, P.,
  \& Cram, L.\ 1983, \solphys, 84, 45

\bibitem[Kudoh et al.(1999)]{Kudoh1999} Kudoh, T., Matsumoto, R.,
  \& Shibata, K. 1999, Comput. Fluid Dynamics J., 8, 56.

\bibitem[Lin et al.(1984)]{Lin1984} Lin, R.~P., Schwartz, R.~A.,
  Kane, S.~R., Pelling, R.~M., \& Hurley, K.~C.\ 1984, \apj, 283, 421

\bibitem[Liu et al.(2004)]{Liu2004} Liu, C., Qiu, J., Gary, D.~E.,
 Krucker, S., \& Wang, H.\ 2004, \apj, 604, 442

\bibitem[Machado et al.(1980)]{Machado1980} Machado M. E., Avrett E. H.,
  Vernazza J. E. et al.\ 1980, \apj, 242, 336

\bibitem[Ning(2008)]{Ning2008} Ning, Z.\ 2008, \apj, 686, 674

\bibitem[Nishizuka et al.(2008)]{Nishizuka2008} Nishizuka, N.,
 Shimizu, M., Nakamura, T., Otsuji, K., Okamoto, T.~J., Katsukawa, Y.,
 \& Shibata, K.\ 2008, \apjl, 683, L83

\bibitem[Petschek(1964)]{Petshek1964} Petschek, H. E., 1964, in
 physics of Solar Flares, ed. W. N. Hess, NASA SP-50, Washington, DC, p. 425

\bibitem[Porter et al.(1984)]{Porter1984} Porter, J.~G., Toomre, J.,
 \& Gebbie, K.~B.\ 1984, \apj, 283, 879

\bibitem[Qiu et al.(2004)]{Qiu2004} Qiu, J., Liu, C., Gary,D.~E.,
  Nita, G.~M., \& Wang, H.\ 2004, \apj, 612, 530

\bibitem[Shimizu et al.(2002)]{Shimizu2002} Shimizu, T., Shine,
 R.~A., Title, A.~M., Tarbell, T.~D., \& Frank, Z.\ 2002, \apj, 574, 1074

\bibitem[Takeuchi \& Shibata(2001)]{Takeuchi2001} Takeuchi, A.,
 \& Shibata, K.\ 2001, \apjl, 546, L73

\bibitem[Tandberg-Hanssen \& Emslie(1988)]{Tand1988}
 Tandberg-Hanssen, E., \& Emslie, A. G. 1988, The physics of
 solar flares, (Cambridge: Cambridge Uni. Press)

\bibitem[Tang et al.(2000)]{Tang2000} Tang, Y.~H., Li, Y.~N.,
 Fang, C., Aulanier, G., Schmieder, B., Demoulin, P.,
 \& Sakurai, T.\ 2000, \apj, 534, 482

\bibitem[Vernazza et al.(1981)]{Vernazza1981} Vernazza, J. E.,
 Avrett, E. H., \& Loeser, R. 1981, \apjs, 45, 635

\bibitem[von Rekowski \& Hood(2008)]{von_Rekowski2008} von Rekowski, B.,
 \& Hood, A.~W.\ 2008, \mnras, 385, 1792

\bibitem[Watanabe et al.(2008)]{Watanabe2008} Watanabe, H., et al.\ 2008,
 \apj, 684, 736

\bibitem[Xia et al.(2007)]{Xia2007} Xia, C., Fang, C.,
  Chen, Q. R., \& Tang, Y. H. 2007, Adv. Space Res., 39, 1402

\bibitem[Yokoyama \& Shibata(2001)]{Yokoyama2001}
 Yokoyama, T., \& Shibata, K.\ 2001, \apj, 549, 1160

\end{thebibliography}
\end{document}